\documentclass[12pt]{article}

\makeatletter \@addtoreset{equation}{section} \makeatother

\usepackage{float}
\usepackage{epsfig,amsfonts}
\usepackage[fleqn]{amsmath}
\usepackage{amsthm,amssymb}
\usepackage{yfonts,mathrsfs}
\usepackage{graphicx}
\usepackage{hhline}
\usepackage{cite}
\def\be{\begin{equation}}
\def\ee{\end{equation}}
\def\bal{\begin{align}}
\def\eal{\end{align}}
\def\bea{\begin{eqnarray}}
\def\eea{\end{eqnarray}}
\def\be{\begin{equation}}
\def\ee{\end{equation}}
\def\bdm{\begin{displaymath}}
\def\edm{\end{displaymath}}
\def\bea{\begin{eqnarray}}
\def\eea{\end{eqnarray}}

\def\sgn{{\rm sgn}}

\def\ri{{\rm i}}
\def\re{{\rm e}}
\def\rd{{\rm d}}

\def\Xint#1{\mathchoice
    {\XXint\displaystyle\textstyle{#1}}%
    {\XXint\textstyle\scriptstyle{#1}}%
    {\XXint\scriptstyle\scriptscriptstyle{#1}}%
    {\XXint\scriptscriptstyle\scriptscriptstyle{#1}}%
    \!\int}
\def\XXint#1#2#3{{\setbox0=\hbox{$#1{#2#3}{\int}$}
    \vcenter{\hbox{$#2#3$}}\kern-.5\wd0}}

\def\dashint{\Xint-}

\begin{document}


\begin{titlepage}

\begin{flushright}
RU-NHETC-2009-01\\
\end{flushright}

\vspace{1.5cm}

\begin{center}
\begin{LARGE}
{\bf On mass spectrum  in 't Hooft's
\\
2D model of mesons}

\end{LARGE}

\vspace{1.3cm}

\begin{large}

{\bf V. A. Fateev$^{1,3}$, S. L. Lukyanov$^{2,3}$ and \\

\vspace{.2cm}

A. B.
Zamolodchikov}$^{2,3}$
\end{large}

\vspace{1.cm}

{${}^{1}$ Laboratoire de Physique Th${\acute {\rm e}}$orique et Astroparticules\\
UMR5207 CNRS-UM2,
Universit${\acute {\rm e}}$ Montpellier II\\
Pl. E. Bataillon, 34095 Montpellier,
France\\

\vspace{.3cm}

${}^{2}$NHETC, Department of Physics and Astronomy\\
     Rutgers University\\
     Piscataway, NJ 08855-0849, USA\\

\vspace{.3cm}
${}^{3}$L.D. Landau Institute for Theoretical Physics\\
  Chernogolovka, 142432, Russia}

\end{center}

\vspace{1.cm}

\begin{center}
\centerline{\bf Abstract} \vspace{.8cm}
\parbox{15.5cm}
{We study 't Hooft's integral equation determining the meson
masses $M_{n}$ in multicolor QCD$_2$. In this note we concentrate
on developing an analytic method, and restrict our attention to
the special case of quark masses $m_1=m_2=g/ \sqrt{\pi}$. Among
our results is systematic large-$n$ expansion, and exact sum rules
for $M_n$. Although we explicitly discuss only the special case,
the method applies to the general case of the quark masses, and we
announce some preliminary results for $m_1=m_2$
(Eqs.\,\eqref{semialpha} and \eqref{gpmalpha}).}
\end{center}

\vspace{0.1cm}

\begin{flushleft}
\rule{4.1 in}{.007 in}\\
{May  2009}
\end{flushleft}
\vfill
\end{titlepage}
\newpage

\section{ Introduction}

As was discovered by G. 't Hooft in 1974 \cite{thooft}, the mass
spectrum of mesons in multi-color QCD in two dimensions admits
for exact solution, because in this model the mesons are
essentially the two-body constructs, and their masses are exactly
determined by the Bethe-Salpeter equation. For the mesons built
from two quarks of  bare (lagrangian) masses $m_1$ and $m_2$, the
Bethe-Salpeter equation reduces to the singular integral equation
\begin{eqnarray}\label{bs0}
2\pi^2\, \lambda\ \ \varphi(x)=
\bigg[\, {{\alpha_1}\over{x}} +
{{\alpha_2}\over{1-x}}\, \bigg]\, \varphi(x) -\  \dashint_{0}^1\rd y\
{{\varphi(y)}\over{(y-x)^2}}\ ,
\end{eqnarray}
where
\begin{eqnarray}\label{alpham}
\alpha_1 = \frac{\pi m_{1}^2}{g^2}-1\,, \qquad \alpha_2 =\frac{\pi
m_{2}^2}{g^2}-1\,,
\end{eqnarray}
with $g$ being the 't Hooft coupling constant (which in $2D$ has the
dimension of  mass). The function $\varphi(x)$ has to
satisfy the boundary conditions
\begin{eqnarray}\label{bc0}
\varphi(0) = \varphi(1) =0\,,
\end{eqnarray}
whence  Eq.\eqref{bs0} defines the spectral problem for the parameter
$\lambda$; the eigenvalues $\lambda_n, n=0,1,2,...$ are
discrete, and determine the meson  masses,
\begin{eqnarray}
M_{n}^2 = 2\pi g^2\ \lambda_n\,.
\end{eqnarray}

In principle, the problem can be solved numerically, to any degree
of accuracy, and over the years a number of approaches were
developed to that end \cite{thooft,Hanson,Huang,Jaffe,Staudacher}.
However, we believe  equation \eqref{bs0} deserves further study
from an analytical standpoint. In our opinion, the most
interesting problem with respect to Eq.\eqref{bs0} is
understanding the analytic properties of the eigenvalues
$\lambda_n$ as the functions of complex $\alpha_1$ and
$\alpha_2$. Without significant analytic input, straightforward
numerical approaches seem to be unsuitable to addressing this
problem.  At the same time, the neat form of equation \eqref{bs0}
suggests that perhaps some analytic information can be extracted.

In this note we report new results about the spectrum
$\{\lambda_n\}$ in the special case\footnote{Note that it is not
the case of massless quarks. In particular, the chiral symmetry
is broken.}
\begin{eqnarray}\label{alpha0}
\alpha_1 = \alpha_2 =0\,.
\end{eqnarray}

Among our results is systematic semiclassical (large-$n$)
expansion of the eigenvalues,
\begin{eqnarray}\label{semiodd}
2\lambda_n =n+{\textstyle\frac{3}{4}}-\frac{2}{3\pi^6\,(n+\frac{3}
{4})^3}+
\frac{2\ (-1)^{n+1}}{\pi^4\,(n+\frac{3}
{4})^2}\, \bigg\{1-
\frac{4\log\big[ \pi \text{e}^{\gamma_E-{1\over
2} }\,(n+ {\frac{3}{4}})\big]}{\pi^2\,
(n+\frac{3}{4})}\,\bigg\}
+O\big({\textstyle{\frac{\log^2(n)}{n^4}}}\big).
\end{eqnarray}
Here
$\gamma_E$ is the Euler constant and we display just three
leading terms, but in principle any number of terms can be
produced via our technique (next four terms can be deduced from
\eqref{jaay},\,\eqref{wkbeven},\,\eqref{wkbodd} and equations
\eqref{lssls},\,\eqref{slsuays} in Appendix A). Note the unusual
logarithmic factors in the third and higher terms, which make
this expansion look rather different from the standard WKB
expansion in the Schr$\ddot {\rm o}$dinger problem. In addition,
our approach allows for analytic evaluation of the spectral sums
\begin{eqnarray}\label{gpm}
G^{(s)}_+ = \sum_{m=0}^{\infty}\,\frac{1}{\lambda_{2m}^{s}}\,,
\qquad\quad G^{(s)}_- =
\sum_{m=0}^{\infty}\,\frac{1}{\lambda_{2m+1}^{s}}
\end{eqnarray}
with integer $s=2,\,3,\,4,\,\ldots$\ . (The sums here are over
even or odd eigenvalues. The corresponding eigenstates are even or
odd with respect to obvious $x\to 1-x$ symmetry of \eqref{bs0}.)
For low $s$ we have, explicitly
\begin{eqnarray}\label{gpmfew}
G^{(2)}_+ &=& 7\,\zeta(3)\, ,\qquad\qquad\qquad\qquad\
 G^{(2)}_- = 2\,,\nonumber\\
G^{(3)}_+ &=& -{\textstyle {4\over 3}}\, \pi^2+28\, \zeta(3)\,,
\qquad\qquad\,
G^{(3)}_- = -{\textstyle {8\over 3}}+{\textstyle {4\over 9}}\, \pi^2\,,\nonumber\\
\\
G^{(4)}_+&=&-2\pi^2 +42\, \zeta(3)-{\textstyle {7\over 3}}\
\pi^2\, \zeta(3) + {\textstyle {49\over 2}}\ \zeta^2(3)+
{\textstyle {31\over 2}}\ \zeta(5)\, ,\nonumber\\
G^{(4)}_-&=&{\textstyle {11\over 3}} -
 {\textstyle {7\over 9}}\ \pi^2+
 {\textstyle {7\over 6}}\ \pi^2\,   \zeta(3) -
{\textstyle {31\over 4}}\ \zeta(5)\nonumber\, .
\end{eqnarray}
Again, in principle analytic expressions for any given $s$ can be
obtained, but for larger $s$ the calculations become increasingly
involved. At the moment we have these numbers up to $s=13$, but
only those with $s=5,\,\ldots\,8$ have sufficiently compact form
to be presented in Appendix A. Put together, the large-$n$
expansion \eqref{semiodd} and the sum rules \eqref{gpm} provide
good control over the entire spectrum: the large-$m$ parts of the
sums \eqref{gpm} can be approximated by the asymptotic expansions
\eqref{semiodd}, thus providing equations for the lower
eigenvalues.

We regard this work as preparatory for studying the spectrum of
\eqref{bs0} with arbitrary $\alpha_1$, $\alpha_2$, with the aim of
understanding analytic properties of the eigenvalues at complex
values of these parameters. We concentrate here on developing the
technique, and the case \eqref{alpha0} is convenient for testing
its efficiency. Besides, many details have particularly neat form
in this case. But for the most part, our technique admits more or
less straightforward extension to the general case, which will be
the next stage of this project. The method also seems to be
suitable for analysis of a large class of Bethe-Salpeter equations
of the type of \eqref{bs0} which emerge in many $2D$ field
theories with confining interactions\footnote{This situation is
typical when one takes a field theory with exact vacuum
degeneracy and adds small interaction which lifts the degeneracy,
giving rise to the confining force between the kinks. The simplest
example is the Ising field theory, in the low-temperature regime,
in the presence of a weak magnetic field \cite{fz2,Rutkevich}.
Unlike the multicolor QCD, where the equation \eqref{bs0} is
exact in the limit $N_c = \infty$, in that case the associated
Bethe-Salpeter equation is only an approximation, expected to be
valid when the magnetic field is sufficiently small, but it seems
to produce meaningful insight into the mass spectrum even at a
large magnetic field.}.

The paper is organized as follows. In Section 2 we discuss general
properties of Eq.\eqref{bs0}. In particular we relate its
solutions to solutions of a certain functional equation (see
Eq.\eqref{fe1}) of the type of Baxter's $T-Q$ equation, with
special analyticity. In Section 3 we develop $\lambda$-series
expansion of the solutions of this equation. This expansion
generates analytic expressions for the spectral sums \eqref{gpm}.
Asymptotic expansion at $\lambda\to\infty$ is developed in Section 4.
It results in the large-$n$ expansion of the eigenvalues
$\lambda_n$. In Section 5 we test these results against the
numerical solution of \eqref{bs0}.

While this paper was in preparation, we have made some progress in
studying the more general case of \eqref{bs0}, with nonzero but
equal values of the parameters
\begin{eqnarray}\label{alfalfa}
\alpha_1 = \alpha_2 = \alpha\,.
\end{eqnarray}
We intend to devote a separate paper to discussing this more
general case, where indeed a very interesting analytic structure
of $\lambda_n (\alpha)$ emerges. But we could not resist the
temptation to announce some results here, which are presented in
Section 6.

\section{Functional equation}

We find it useful to recast  Eq.\eqref{bs0} into somewhat
different form, via the integral transformation
\begin{eqnarray}\label{klals}
\varphi(x)=\int_{-\infty}^{\infty}{\rd\nu\over 2\pi}\  \Psi(\nu)\
 \big({\textstyle{x\over 1-x}}\big)^{\frac{\ri\nu}{2}}\, ,
\quad
\Psi(\nu)=\int_{0}^{1}{\rd x\over 2x(1-x)} \ \ \varphi(x)\
\big({\textstyle{x\over 1-x}}\big)^{-\frac{\ri\nu}{2}}\, ,
\end{eqnarray}
which is just  Fourier transform with respect to the variable
$\frac{1}{2}\,\log(\frac{x}{1-x})$ (This transformation was
previously used in Ref.\cite{neuberger}). The $\nu$-space form of
\eqref{bs0} is
\begin{eqnarray}\label{bsnu}
\nu\,\coth\big({\textstyle\frac{\pi\nu}{2}}\big)\ \Psi(\nu) -
\lambda\,\int_{-\infty}^{\infty}\rd\nu'\ S(\nu-\nu')\ \Psi(\nu')=0\, ,
\end{eqnarray}
where the kernel
\begin{eqnarray}\label{alsksa}
S(\nu) = \frac{\pi\nu}{2\,\sinh(\frac{\pi\nu}{2})}
\end{eqnarray}
in the right hand side is regular at all real $\nu$. The solution
$\Psi(\nu)$ must decay at $|\nu|\to\infty$ (for the norm
$\parallel\varphi\parallel^2=\int_{0}^1\rd x\  |\varphi(x)|^2$ to
be finite), and it must be a smooth function of $\nu$ (for the
function $\varphi(x)$ in \eqref{bs0} to satisfy the boundary
conditions \eqref{bc0}). The spectrum $\{\lambda_n\}$ is
determined by the existence of solutions which satisfy these
conditions. In fact, both these conditions, once satisfied,  are
satisfied with substantial redundancy.

Equation \eqref{bsnu} dictates that any smooth solution is in
fact analytic. Moreover, it is possible to show that the solutions
$\Psi(\nu)$ are meromorphic functions of $\nu$, with the poles at
$\nu = \pm (2 k-1)\, \ri$, $k\in \mathbb{Z}$, of the order $k\in
{\mathbb N}$. In particular, the function $Q(\nu)$ defined as
\begin{eqnarray}\label{psiq}
Q(\nu) = \nu\,\cosh\big({\textstyle\frac{\pi\nu}{2}}\big)\ \Psi(\nu)
\end{eqnarray}
is analytic in the strip $|\Im m\,\nu| \leq 2$, grows slower then
any exponential of $\nu$ at infinity, and turns to zero at $\nu=0,\,
\pm 2\ri$, i.e.
\begin{eqnarray}\label{conditions}
Q(0) = Q(\pm 2\ri) =0\,.
\end{eqnarray}
Under these conditions the integral operator in the right hand
side can be inverted in terms of finite difference operator
(which is derived by standard manipulations with  shifts of
the integration contour), leading to the functional equation
\begin{eqnarray}\label{fe1}
Q(\nu+2\ri) + Q(\nu-2\ri)-2\,Q(\nu) = -4\pi\lambda\ \ \nu^{-1}\,
\tanh\big({\textstyle\frac{\pi\nu}{2}}\big)\ Q(\nu)\,.
\end{eqnarray}
Equation \eqref{fe1} is the basis of our analysis below.

A quick look at the asymptotic form of \eqref{fe1} at $\Re e\, \nu
\to \infty$ reveals that its solutions generally behave as $\re^{
k\,\nu}\,f(\nu)$, with integer $k$, and $f(\nu)$ bounded by any
exponential. Obviously, any positive $k$ would violate the
asymptotic condition for $\Psi(\nu)$. Thus, we are interested in
the solutions which are bounded as
\begin{eqnarray}
Q(\nu) = O\big(\,\re^{\epsilon |\nu|}\,\big)\qquad \text{as}\quad |\Re e\,\nu| \to
\infty\,,
\end{eqnarray}
with any $\epsilon$. Note that this condition implies that the function
$\Psi(\nu)$ in fact decays exponentially in this limit.

A solution of \eqref{fe1} with the desired analytic and
asymptotic properties exists only at specific values of
$\lambda$, which determine the eigenvalues of \eqref{bsnu}.
However, if the conditions \eqref{conditions} are relaxed, the
solutions $Q(\nu|\lambda)$ exist at any $\lambda$. For generic
$\lambda$, the associated function $\Psi(\nu|\lambda)$ no longer
satisfies the integral equation \eqref{bsnu}. Instead, it solves
the related inhomogeneous equation
\begin{eqnarray}\label{unh0}
\nu\,\coth\big({\textstyle\frac{\pi\nu}{2}}\big)\ \Psi(\nu|\lambda) -
\lambda\ \dashint_{-\infty}^{\infty}\rd\nu'\ S(\nu-\nu')\ \Psi(\nu'|\lambda)
= F(\nu|\lambda)\,,
\end{eqnarray}
where
\begin{eqnarray}
F(\nu|\lambda) = \frac{q_{+}(\lambda)\,\nu
+q_{-}(\lambda)}{\sinh({\pi\nu\over 2})}\,,
\end{eqnarray}
with the coefficients $q_{+}(\lambda)$ and $q_{-}(\lambda)$
related to the values $Q(0|\lambda)$ and $Q(\pm 2\ri|\lambda)$ in
a linear manner (note that in view of the functional equation
\eqref{fe1}, only two of these values are independent). Since now
in general $Q(0|\lambda)\neq 0$, the integrand in the l.h.s.
involves a first order pole at $\nu'=0$, and the integral is
understood as its principal value. It is possible to show that,
given the coefficients $q_\pm$, the solution of \eqref{unh0} is
unique. These coefficients can be chosen at will, and therefore
the equation \eqref{unh0} generates two-dimensional space of
functions $\Psi(\nu|\lambda)$. It is natural to choose the basis
in accord with the obvious $\nu\to-\nu$ symmetry of the problem.
We thus define symmetric and antisymmetric basic functions,
\begin{eqnarray}\label{psisym}
\Psi_{\pm}(-\nu|\lambda) = \pm\,\Psi_{\pm}(\nu|\lambda)\,,
\end{eqnarray}
which solve the equation \eqref{unh0} with $F(\nu|\lambda)$ in
the r.h.s. taken to be
\begin{eqnarray}\label{fpm}
F_{+}(\nu) = \frac{\nu}{\sinh(\frac{\pi\nu}{2})}\qquad
\text{and}\qquad F_{-}(\nu) = \frac{1}{\sinh(\frac{\pi\nu}{2})}\ ,
\end{eqnarray}
respectively.

At the spectral points $\lambda=\lambda_n$ the original equation
\eqref{bsnu} is to be recovered. That means that at certain values
of $\lambda$ the basic functions $\Psi_{\pm}(\nu|\lambda)$
diverge. More precisely, Eq.\,\eqref{unh0} can be rewritten in the form of  an inhomogeneous
Fredholm integral equation of the second kind (see Appendix B for details)
and  it follows
from the resolvent formalism that $\Psi_{\pm}(\nu|\lambda)$ are
meromorphic functions of $\lambda$, with only poles at the
eigenvalues of \eqref{bsnu},
\begin{eqnarray}\label{lambdapoles}
\Psi_{+}(\nu|\lambda)= \sum_{m=0}^{\infty}
\frac{{\rm c}_{2m}\Psi_{2m}(\nu)}{\lambda-\lambda_{2m}}\,,
\quad\quad \Psi_{-}(\nu|\lambda)=
\sum_{m=0}^{\infty}\ \frac{{\rm
c}_{2m+1}\Psi_{2m+1}(\nu)}{\lambda-\lambda_{2m+1}}\ ,
\end{eqnarray}
where, as we have mentioned in  Introduction, $\lambda_{2m}$ and
$\lambda_{2m+1}$, $m=0,\,1,\,2,\ldots$, refer to the eigenvalues
of \eqref{bsnu} in the even and odd sectors, respectively, and
$\Psi_{2m}(\nu)$ and $\Psi_{2m+1}(\nu)$ are associated
eigenfunctions\footnote{Here and below $\Psi_n(\nu)$ stand for
normalized eigenfunctions, i.e. we assume that $\int_0^1\rd x\,
|\varphi_n(x)|^2=1$ for the associated $\varphi_n(x)$,
Eq.\eqref{klals}.}.

It is useful to note that the functions $\Psi_{+}(\nu|\lambda)$
and $\Psi_{-}(\nu|\lambda)$ are related to
the ``quark form factors''  of the vector
current $J_\mu = {\bar\psi}\gamma_\mu \psi$ and
the scalar density $S={\bar\psi}\psi$,
respectively, with the parameter $\lambda$ (more precisely
$2\pi\,g^2\,\lambda$) interpreted as $q^2$, the square of the
total 2-momentum (see Refs.\cite{gross, einhorn}, where inhomogeneous
integral equations equivalent to \eqref{unh0}, \eqref{fpm} appear in this
connection). Therefore the structure \eqref{lambdapoles} is
well expected, and the coefficients ${\rm c}_n$ in
\eqref{lambdapoles} are related  to the matrix elements
\begin{eqnarray}
\langle\,0 \mid\,J_{\mu}(0)\,\mid M_{2m}, q\,\rangle &=& \ri\
\epsilon_{\mu\nu}\,q^\nu\ \sqrt{N_c}\ \pi^{3\over 2}\ \ {\rm c}_{2m}\ ,\\
\langle\,0\mid S(0) \mid M_{2m+1}, q\,\rangle
&=& 2\pi g\ \ \sqrt{N_c}\  \  {\rm c}_{2m+1}\ ,\nonumber
\end{eqnarray}
where $\mid  M_n, q \,\rangle$ stands for the $n$-th meson state
with 2-momentum $q$. Let us mention here the neat expressions for
the current-current correlation function in terms of
$\Psi_{+}(\nu|\lambda)$,
\begin{eqnarray}
\langle\,J_{\mu}(q)\,J_{\nu}(-q)\,\rangle = {\ri N_c\over \pi}\
\Big(\,\frac{q_\mu q_\nu} {q^2}- g_{\mu\nu} \Big)\ \big[\,1
-\Psi_{+}(0|\lambda)\, \big]\ .
\end{eqnarray}

Having in mind this analyticity in $\lambda$, our strategy in
solving the problem will be as follows. Starting with equation
\eqref{fe1}, we will be looking for two solutions,
$Q_{+}(\nu|\lambda)$ and $Q_{-}(\nu|\lambda)$ of the functional
equation \eqref{fe1}, analytic in the strip $|\Im m \,\nu|\leq
2$, and growing slower than any exponential at $|\Re e \,\nu| \to
\infty$. We will assume that
\begin{eqnarray}\label{qsym}
Q_{\pm}(-\nu|\lambda) = \mp\,Q_{\pm}(\nu|\lambda)\,,
\end{eqnarray}
and fix the normalizations by the conditions
\begin{eqnarray}\label{qnorm}
Q_{+}(2\ri|\lambda)=-Q_{+}(-2\ri|\lambda)=2\ri\,, \qquad
Q_{-}(0|\lambda)=1\,.
\end{eqnarray}
Then the functions $\Psi_{\pm}(\nu|\lambda)$ related to
$Q_{\pm}(\nu|\lambda)$ as in \eqref{psiq} have appropriate
symmetry \eqref{psisym}, and solve the equation \eqref{unh0}
precisely with the right-hand sides \eqref{fpm}, as one can
readily verify. In fact, the remarkably simple formula
\begin{eqnarray}\label{rf}
\partial_\lambda \log D_{\pm}(\lambda) = 2\ri\,\partial_\nu\  \log
Q_{\mp}(\nu|\lambda)\big|_{\nu=\ri}
\end{eqnarray}
relates the logarithmic derivatives of $Q_\mp$ at $\nu=\ri$
to  suitably defined spectral determinants,
\begin{eqnarray}\label{dpm}
D_{+}(\lambda) =
\prod_{m=0}^{\infty}\bigg(1-\frac{\lambda}{\lambda_{2m}}\bigg) \
\re^\frac{\lambda}{\lambda_{2m}}\,, \qquad D_{-}(\lambda) =
\re^{2\lambda}\,\prod_{m=0}^{\infty}\bigg(1-\frac{\lambda}{\lambda_{2m+1}}\bigg)
\ \re^\frac{\lambda}{\lambda_{2m+1}}\,.
\end{eqnarray}
(To be precise, somewhat more complicated expressions,
Eqs.\,\eqref{kkssa}, follow directly from the integral equation\
\eqref{unh0}. See Appendix B, where we explain the status of
Eq.\,\eqref{rf}.) Note that this relation is insensitive to
normalization conditions assumed for $Q_{\pm}(\nu|\lambda)$. In
what follows, we develop two different expansions for such
solutions $Q_{\pm}(\nu|\lambda)$. One is just the power series in
$\lambda$, and the other is the asymptotic expansion around the
essential singularity $\lambda=\infty$. Eq.\eqref{rf} translates
the former expansions into the sum rules \eqref{gpm}, while the
latter ones lead to the large-$n$ expansion \eqref{semiodd}.

Before turning to the details, let us make the following remark.
The functional equation \eqref{fe1} has the form of the famous
$T-Q$ relation of Baxter, and many general statements can be
adopted to our case. In particular, it is easy to show that the
so-called quantum Wronskian built from the two solutions
$Q_{\pm}(\nu|\lambda)$ is a constant,
\begin{eqnarray}\label{qw}
Q_{+}(\nu+\ri | \lambda)\,Q_{-}(\nu-\ri|\lambda) -
Q_{+}(\nu-\ri|\lambda)\, Q_{-}(\nu+\ri|\lambda) =2\ri\ .
\end{eqnarray}
The fact that this combination does not depend on $\nu$ follows
from the functional equation \eqref{fe1}, and the asymptotic
conditions at $|\Re e\,\nu|\to\infty$. The particular value $2\ri$
in the r.h.s. reflects the special normalization \eqref{qnorm};
with different normalization it would be a different (generally
$\lambda$-dependent) constant. This equation, combined with
\eqref{rf}, allows one to establish some useful relations. It
follows from \eqref{rf} that $Q_{+}(\ri|\lambda)$ turns to zero at
the odd spectral values $\lambda=\lambda_{2m+1}$ (likewise,
$Q_{-}(\ri|\lambda)$ does the same at the even values
$\lambda_{2m}$).  But the identity $Q_{+}(\ri|\lambda)\,
Q_{-}(\ri|\lambda)=\ri $ (elementary consequence of \eqref{qw})
shows that $\lambda_{2m+1}$ exhaust all zeros of
$Q_{+}(\ri|\lambda)$ viewed as the function of $\lambda$. In other
words, $Q_{+}(\ri|\lambda)$ must be proportional to
$D_{-}(\lambda)/D_{+}(\lambda)$, up to a factor which is entire
function of $\lambda$ with no zeros, i.e. in our case the factor
of the form $\exp(a+b\lambda)$. More careful analysis (in the next
section) allows one to fix this ambiguity completely. Let us
present here the result  in the form
\begin{eqnarray}\label{qdrel}
\frac{Q_{+}(\ri|\lambda)}{Q_{+}(2\ri|\lambda)}={{1\over 2}}\ \frac{D_{-}(\lambda)}
{D_{+}(\lambda)}\,, \quad\ \ \  \
\frac{Q_{-}(\ri|\lambda)}{Q_{-}(0|\lambda)}=2\ri\ \frac{D_{+}(\lambda)}
{D_{-}(\lambda)}\,,
\end{eqnarray}
insensitive to normalizations of $Q_{\pm}(\nu|\lambda)$.

\section{Expansion in powers of $\lambda$}

In principle, one can generate the expansion in $\lambda$ just by
iterating the integral equation \eqref{unh0}, with the right-hand
side taken in one of the forms \eqref{fpm}
(see Eqs.\,\eqref{lsaaaiia} in Appendix B).
This leads to the convergent series
\begin{eqnarray}\label{lambdaser}
Q_{\pm}(\nu|\lambda) = \sum_{s=0}^{\infty} Q_{\pm}^{(s)}(\nu)\
\lambda^s\,,
\end{eqnarray}
with the coefficients given by $s$-fold integrals involving the
kernel $S(\nu)$. Direct evaluation of these integrals is
difficult, and therefore we take another approach based on the
functional equation \eqref{fe1}. We look for the solution of
\eqref{fe1} in the form of the power series \eqref{lambdaser},
with the coefficients $Q_{\pm}^{(s)}(\nu)$ having the symmetry
\eqref{qsym}, analytic in the strip $|\Im m \,\nu| \leq 2$, and
growing slower than any exponential at $|\Re e\,\nu|\to\infty$. It
is clear upfront that at each order these conditions fix the
coefficients uniquely (after all, it is just a somewhat indirect
way of evaluating the integrals appearing in the iterative
solution of \eqref{unh0}). But the solution for
$Q_{\pm}^{(s)}(\nu)$ obtained this way involves polynomials of
$\nu$ of growing degree, and the expressions quickly become
cumbersome. The following observation greatly facilitates the
calculations.

Note that the factor
\begin{eqnarray}\label{zdef}
z \equiv 2\pi\lambda\,\tanh\big({\textstyle\frac{\pi\nu}{2}}\big)
\end{eqnarray}
in the r.h.s. of this equation, is insensitive to the shifts $\nu
\to \nu \pm 2\ri$, and if no attention to the analytic properties
is paid, it can be regarded as constant. Then the equation
\eqref{fe1}, written as
\begin{eqnarray}\label{fe2}
Q(\nu+2\ri) + Q(\nu-2\ri)-2\,Q(\nu) =
-2z\ \nu^{-1}\, Q(\nu)
\end{eqnarray}
is recognizable as one of the recursion relations satisfied by
confluent hypergeometric functions \cite{Abramowitz}. Specifically, the
functions $\nu\,M(1+{\ri \nu\over 2},2,- \ri z)$ and
$\Gamma(1+{\ri \nu\over 2})\,U(1+{\ri \nu\over 2},2,- \ri z)$
are known to satisfy
\eqref{fe2} (see e.g. \cite{Abramowitz}, Eqs.\,13.4.1,\,13.4.15). Here the conventional
notations $M(a,c,x)$ and $U(a,c,x)$ for two canonical solutions of
confluent hypergeometric equation are used. Of course, by
themselves these functions do not provide solution to our
problem, since they have wrong analyticity in $\nu$. For one, the
second of these functions has logarithmic singularity at $z=0$,
which in view of \eqref{zdef} produces unpleasant branching
points in the $\nu$-plane. This problem is easy to cure by
observing that the logarithmic term by itself satisfies
Eq.\eqref{fe2}, and subtracting it produces another solution
which is now a single-valued function of $\nu$. Thus, we found it
convenient to use the combinations
\begin{eqnarray}\label{up}
M_{+}(\nu,z) &=& \nu\,\re^{\ri z\over 2}\ M\big(1+{\textstyle\frac{\ri\nu}{2}},\,2, - \ri z \big)\, ,\\
M_{-}(\nu,z)& =&  -\ri z\,
\re^{\ri z\over 2}\,\Gamma\big(1+{\textstyle \frac{\ri\nu}{2}}\big)\,
U\big(1+{\textstyle\frac{\ri\nu}{2}},2, - \ri z \big)-\nonumber\\
&&{\textstyle{1\over 2}}\,
 \big[\,  z\,\log\big( -{\textstyle{\ri\over 4}}\, z\,\re^{\gamma_E}\,
\big)+\ri\pi^2\lambda\,
\big]\ {M}_{+}(\nu,z)\ ,\label{um}
\end{eqnarray}
where the coefficients $\re^{\ri z\over 2}$ and the extra constant in the
brackets in \eqref{um} are chosen to ensure the symmetry
\begin{eqnarray}
M_{\pm}(-\nu,-z) = \mp M_{\pm}(\nu,z)\,,
\end{eqnarray}
in accord with the obvious symmetry of  Eq.\eqref{fe2}. Both
\eqref{up} and \eqref{um} are entire functions of $z$, in
particular both can be represented by convergent expansions in
the powers of $z$:

\begin{eqnarray}\label{sslksa}
M_+(\nu, z)&=&\nu\,\re^{\ri z\over 2}\
\sum_{s=0}^\infty {(1+{\ri\nu\over 2})_s\over (s+1)! s!}\ \ (-\ri
z)^s\ ,
\\
M_-(\nu, z)&=& {\textstyle{1\over 2}}\ \Big[\ \re^{\ri z\over
2}\, \Sigma(\nu,z)+ \re^{-{\ri z\over 2}}\, \Sigma(-\nu,-z)\
\Big]\, ,\nonumber \eea where \bea\label{sjsak} \Sigma(\nu, z)=1+
\sum_{s=1}^\infty {({\ri\nu\over 2})_{s}\over s! (s-1)!}\  \Big[\,
\psi\big(s+{\textstyle{\ri\nu\over 2}}\big)-\psi(s)-\psi(s+1)+
\psi\big({\textstyle {1\over 2}}\big) \,\Big]\, \ (-\ri z)^{s}\ .
\end{eqnarray}
These expansions make explicit a more serious problem. In view
of \eqref{zdef}, each term of this expansion produces poles at
$\nu=\pm\,\ri$, of growing order, and thus both \eqref{up} and
\eqref{um}, viewed as the functions of $\nu$ at fixed $\lambda$,
have essential singularities at these points, whereas we need
solutions of \eqref{fe1} analytic in the strip $|\Im m\,\nu|\leq
2$. We are thus compelled to look for the solutions in the form
\begin{eqnarray}\label{gsol}
&&Q_{\pm}(\nu|\lambda) =
A_{\pm}(z,\lambda)\ M_{\pm}(\nu,z)+B_{\pm}(z,\lambda)\ z\, M_{\mp}(\nu,z)\ ,
\end{eqnarray}
where the coefficients, entire functions of $z^2$, are to be
adjusted to compensate for the above singularity at $z=\infty$. So
far we were unable to find the closed form solution of this
analytic problem. But it is easy to generate the solution as an
expansion in the powers of $\lambda$. In view of the above
analyticity, we assume that the coefficients can be expanded in
double series in $\lambda$ and $z^2$. Regarding them as the
functions of $\nu$ and $\lambda$ (through the relation
\eqref{zdef}), this expansions has the form of the power series
\begin{eqnarray}
A_{\pm}(z,\lambda) = \sum_{s=0}^{\infty}\,a_{\pm}^{(s)}
(\tau)\,\lambda^s\,,\qquad\qquad  B_{\pm}(z,\lambda)
=\sum_{s=0}^{\infty}\,b_{\pm}^{(s)} (\tau)\,\lambda^s
\end{eqnarray}
with $a_{\pm}^{(s)}(\tau)$ and $b_{\pm}^{(s)}(\tau)$ being
polynomials in $\tau \equiv({z\over 4\lambda})^2= {\pi^2\over 4}
\tanh^2(\frac{\pi\nu}{2})$, of the highest degree $[s/2]$. The
numerical coefficients in these polynomials are to be adjusted in
such a way as to compensate for all the pole terms generated by
the expansions of the functions $M_{\pm}(\nu,z)$ in \eqref{gsol},
order by order in $\lambda$. The remaining constant terms are
then fixed by the normalization conditions \eqref{qnorm} which
demand that $A_\pm(0,\lambda)=1$. Clearly, this linear problem at
each order has a unique solution. We have calculated explicitly
the polynomials $a_{\pm}^{(s)}(\tau)$, $b_{\pm}^{(s)}(\tau)$ up to
$s=13$. Let us present here the first few of them, just to give
the flavor of it:
\begin{eqnarray} && a^{(2)}_+=\tau\,
,\qquad\qquad a^{(3)}_+={\textstyle{64\over 9}}\,\tau\, , \qquad
a^{(4)}_+=
{\textstyle{1\over 4}}\ \tau\ \big(\,50+ 21\,\zeta(3)-5\tau\, \big)\, ,\nonumber \\
&&b_+^{(0)}={\textstyle{1\over 2}}\, , \qquad\qquad    b^{(1)}_-={\textstyle{4\over 3}}\, ,
\qquad\ \  \ \  b^{(2)}_-={\textstyle{1\over 4}}\ \big(\,6+7\,\zeta(3)-\tau\, \big)\, ,
\end{eqnarray}
and
\begin{eqnarray}
&&a^{(2)}_-=-\tau\, ,\qquad\
a^{(3)}_-={\textstyle{8\over 9}}\,\tau\, , \qquad \ \ \   a^{(4)}_-={\textstyle{1\over 12}}\ \tau\
\big(\, 21\,\zeta(3)-14+3\tau\,\big)\, , \nonumber \\
&&b_-^{(0)}=0\, ,\qquad\  \ \ \    b^{(1)}_-=4\, ,\qquad\qquad
b^{(2)}_-={\textstyle{7\over 2}}\, \zeta(3)-5-\tau\, .
\end{eqnarray}

Eq.\eqref{rf} makes it straightforward to convert the
$\lambda$-expansions of $Q_{\pm}(\nu|\lambda)$ into the
expansions of the spectral determinants \eqref{dpm},
\begin{eqnarray}
\label{slslksasa}
\log D_{\pm}(\lambda) =(1\mp 1)\, \lambda- \sum_{s=2}^{\infty}\,
s^{-1}\ G_{\pm}^{(s)}\ \lambda^s\,,
\end{eqnarray}
where the coefficients give explicitly the spectral sums
\eqref{gpm}. For the few lowest $s$ the result of this
calculation was already displayed in \eqref{gpmfew}, but we
present many more in  Appendix A. With many $G^{(\pm)}_{s}$
known, the sum rules \eqref{gpm} become a useful tool in
determining the eigenvalues $\lambda_n$, especially so when
combined with the large-$n$ asymptotic expansions, which we derive
in the next section.

\section{Asymptotic expansion at $\lambda\to\infty$}

To develop the large-$\lambda$ expansions of the functions
$Q_{\pm}(\nu|\lambda)$ we start by constructing a formal solution
of the functional equation \eqref{fe1}, of the following structure
\begin{eqnarray}\label{sform}
S(\nu|\lambda) =(-\lambda)^{-\frac{\ri\nu}{2}}\
\sum_{k=0}^{\infty}\,S_{k}(\nu)\,{\lambda^{-k}}\,.
\end{eqnarray}
It is impossible to satisfy all the analytic conditions required
for the functions $Q_{\pm}(\nu|\lambda)$ within this ansatz, but
we would like to get as close to the desired analyticity as
possible. In particular, we demand that the coefficients $S_k
(\nu)$ are meromorphic functions of $\nu$, growing slower then any
exponential at $|\Re e\,\nu| \to \infty$. The form \eqref{sform}
is obviously designed to serve the case of negative real
$\lambda$, if we choose the principal branch of
$(-\lambda)^{-\frac{\ri\nu}{2}}$ (other branches exhibit
unacceptable exponential growth at $|\Re e\,\nu|\to\infty$).

Plugging this expansion into \eqref{fe1} generates a sequence of
recurrent functional equations for the coefficient functions
$S_k(\nu)$. In the zeros order we have
\begin{eqnarray}\label{rec}
S_{0}(\nu+2\ri)={\frac{4\pi}{\nu}}\,\tanh\Big({\frac{\pi\nu}{2}}\Big)\
S_{0}(\nu)\,.
\end{eqnarray}
The solution of this equation, analytic in the strip $|\Im m
\,\nu|\leq 2$ and bounded at $|\Re e\,\nu| \to \infty$, is unique
up to a normalization. It can be written in an explicit
form\footnote{The function $\psi_0 (\nu) = \frac{1}{\nu}\,
\tanh\left(\frac{\pi\nu}{2}\right)\,S_0(\nu)$, with $S_0(\nu)$ as
in \eqref{w0}, provides an exact solution to the ``scattering''
problem
\begin{eqnarray}\nonumber
-\,\varphi(x)=\dashint_{0}^{\infty}\,\frac{\varphi(y)}{(x-y)^2}\,dy
\end{eqnarray}
associated with \eqref{bs0}, see Ref.\cite{Brauer}. Namely,
\begin{eqnarray}\nonumber
\varphi(x) =
\int_{-\infty}^{\infty}\rd\nu\ x^{-\frac{\ri\nu}{2}}\,\,\psi_{0}(\nu)\ .
\end{eqnarray}
Using the known asymptotic behavior of the Barnes  $G$-function
\cite{Barns}, it
is straightforward to derive the ``scattering phase'' in
\begin{eqnarray}\nonumber
\varphi(x) \to \re^{\frac{3\pi \ri}{8}}\ \re^{-\ri x}+\re^{-\frac{3\pi
\ri}{8}}\ \re^{\ri x}\quad \text{as} \quad x\to\infty\ .
\end{eqnarray}
from which the constant term $\frac{3}{4}$ in
Eq.\eqref{semiodd} (already conjectured in
Ref.\cite{thooft}) follows. Our analysis in this section goes
beyond this simple approximation.}
\begin{eqnarray}\label{w0}
S_{0}(\nu) = \left(2\pi\right)^{-{1\over 2}-\frac{\ri\nu}{2}}\,\
\frac{G\left(2+\frac{\ri\nu}{2}\right)\,G\left(\frac{1}{2}-\frac{\ri\nu}{2}\right)}
{G\left(1-\frac{\ri\nu}{2}\right)\,G\left(\frac{3}{2}+\frac{\ri\nu}{2}\right)}\ \ \ \ \ \ \ \ \ \ \ \
\big(\, S_0(\ri)=1\,\big)
\end{eqnarray}
in terms of the Barnes  $G$-function  (see e.g.
\cite{Barns})
\begin{eqnarray}
G(x+1)=(2\pi)^{x\over 2}\ \re^{-{x(x+1)\over 2}-{\gamma_E\over 2}\, x^2}\
\prod_{n=1}^{\infty}\bigg[\,
\Big(1+{x \over n}\Big)^{n}\, \re^{-x+{x^2\over 2n}}\,\bigg]\ .
\end{eqnarray}

At higher orders in $\lambda^{-1}$ the equation \eqref{fe1} leads
to the recurrent relations of the form
\begin{eqnarray}\label{sigmaeq}
\sigma_k (\nu+2\ri)-\sigma_k(\nu) = \rho_k(\nu)
\end{eqnarray}
for the ratios $\sigma_k(\nu)=S_k(\nu)/S_0(\nu)$, with
$\rho_k(\nu)$ being certain expressions involving $\sigma_{k'}
(\nu)$ from the lower orders $k'=k-1,\,k-2$. While beyond the leading
order no solutions analytic in the strip $|\Im m\,\nu|\leq 2$
exist, it is possible to find solutions analytic in that strip
except for the points $\nu=0, \pm 2\ri$, where the poles of growing
order appear. The result of this calculation is summarized by the
formula
\begin{eqnarray}\label{ssol}
S(\nu|\lambda) = R(z,\lambda)\,{\hat U}(\nu,z)\,,
\end{eqnarray}
where ${\hat U}(\nu,z)$ stands for the formal  asymptotic series
\begin{eqnarray}\label{uhat}
{\hat U}(\nu,z) = (-\lambda)^{-\frac{\ri\nu}{2}}\ S_0 (\nu)\
\sum_{k=0}^{\infty}\,\frac{\left(1+\frac{\ri\nu}{2}\right)_k
\,\left(\frac{\ri\nu}{2}\right)_k}{k!}\ (\,\ri z\,)^{-k}\,,
\end{eqnarray}
and $z$ is the same combination \eqref{zdef}, insensitive to the
shifts $\nu \to \nu\pm 2\ri$. The fact that \eqref{uhat} satisfies
\eqref{fe1} can be verified directly, but it is clear upfront
from the following observation; The series appearing in
\eqref{uhat}, when  multiplied by $\Gamma(1+{\ri\nu\over 2})\
(-\ri z)^{-1-{\ri\nu\over 2}}$, coincides with the asymptotic
expansion of the  function $\Gamma(1+{\ri\nu\over 2})\
U(1+{\ri\nu\over 2},2,-\ri z)$, which  satisfies \eqref{fe1}. In
writing \eqref{uhat} we simply replaced the overall factor
$\Gamma(1+{\ri\nu\over 2})\ (-\ri z)^{-1-\frac{\ri\nu}{2}}$ by
the much more analytically attractive
$(-\lambda)^{-\frac{\ri\nu}{2}}\ S_0 (\nu)$. The factor
$R(z,\lambda)$ in \eqref{ssol} represents the ambiguities in the
solutions of \eqref{sigmaeq}; it is to be understood as a formal
series in the powers of $z^{-1}$ and $\lambda^{-1}$, or
equivalently as a series in $\lambda^{-1}$ with the coefficients
being polynomials in the variable
\begin{eqnarray}\label{cot}
c\equiv\, \ri\pi\, \coth\big({\textstyle\frac{\pi\nu}{2}}\big)\,.
\end{eqnarray}

The asymptotic expansions of the functions $Q_{\pm}(\nu|\lambda)$
can be built from the formal solution \eqref{uhat} in much the
same way as the $\lambda$-expansions were constructed from the
basic functions \eqref{up}, \eqref{um} in the previous section.
Having in mind the symmetry \eqref{qsym}, we look for
$Q_{\pm}(\nu|\lambda)$ in the form \footnote{Here and below the
symbol $\asymp$ stands for equality in the sense of asymptotic
series.}
\begin{eqnarray}\label{qpmass}
Q_{\pm}(\nu|\lambda)\asymp R_{\pm}(z,\lambda)\,{\hat
U}(\nu|z)\,\mp\,R_{\pm}(-z,\lambda)\,{\hat U}(-\nu|-z)\,.
\end{eqnarray}
The coefficients $R_{\pm}(z,\lambda)$ are to be adjusted to fix
the analytic problems present in ${\hat U}(\nu|z)$ and ${\hat
U}(-\nu|-z)$.
One of these problems was already mentioned above. The series
\eqref{uhat} explicitly exhibits at each order in $\lambda^{-1}$
poles at $\nu=0,\pm 2\ri$, of the growing order.  This problem can
be fixed order by order in $\lambda^{-1}$, with
$R_{\pm}(z,\lambda)$ taken in the form
\begin{eqnarray}\label{rexp}
R_{\pm}(z,\lambda)\propto 1+\sum_{k=1}^{\infty}\,R_{\pm}^{(k)}(c,L)\,\lambda^{-k}\,,
\end{eqnarray}
with $R_{\pm}^{(k)}(c,L)$, polynomials in the cotangent
\eqref{cot}, adjusted to cancel these poles. Because of the
factor $(-\lambda)^{-\frac{\ri\nu}{2}}$, the Laurent expansions of
\eqref{uhat} around the poles generate logarithms of $-\lambda$.
As a result, the coefficients $R_{\pm}^{(k)}$ in \eqref{rexp}
emerging in this calculation turn out to be also polynomials in
the variable
\begin{eqnarray}\label{Ldef}
L=\log\left(-2\pi\lambda\right)+\gamma_E\,.
\end{eqnarray}
This is a novel feature of the large-$\lambda$ expansion, which
ultimately leads to the logarithmic factors in the expansions
\eqref{semiodd}. As expected, the solution of this
pole-cancellation problem at each order in $\lambda^{-1}$ turns
out to be essentially unique, i.e. unique up to terms which can be
absorbed into the overall normalization of $Q_{\pm}(\nu|\lambda)$.
With the relations \eqref{qdrel}, the normalization conditions
\eqref{qnorm} imply the following general form of  the
coefficients $R_{\pm}(z,\lambda)$:
\begin{eqnarray}\label{ksaksas}
R_{\pm}(z,\lambda)=(-\lambda)^{-{1\over 2}}\  \ri^{1\mp 1\over 2}\
\bigg[\, {D_-(\lambda)\over 2D_+(\lambda)}\, \bigg]^{\pm 1} \
\bigg[\, 1+c\,
 \sum_{k=1}^\infty  P^{(k)}_{\pm}(c,L)\
\lambda^{-k}\, \bigg]\ .
\end{eqnarray}
We have explicitly computed the polynomials $P_{\pm}^{(k)}(c,L)$
up to $k=7$, but display here only the first few of them (again,
just to give the flavor of the emerging expressions):
\begin{eqnarray}\label{pk}
P_{\pm}^{(1)}(c,L)=0\,, \qquad  P_{\pm}^{(2)}(c,L)={{\pm 1\over
4\pi^4}} \,, \qquad P_{\pm}^{(3)}(c,L)={{\pm 1\over 24\pi^6}}\ (
6\,  c-12 L+6\mp 1)\ .
\end{eqnarray}
The overall factors in \eqref{ksaksas} are related to the ratio of
the spectral determinants\ \eqref{dpm} via \eqref{qdrel}; the expansion
\bea\label{slsssosil}
{D_-(\lambda)\over D_+(\lambda)}\asymp \sqrt{2\over -\lambda\pi^2}\
\exp\bigg[\,{1\over 2\lambda\pi^2}
-{L \over 2(\lambda\pi^2)^2}+
{6 L(L-1)-\pi^2-1\over 12 (\lambda\pi^2)^3}+O(L^3\lambda^{-4})\, \bigg]
\eea
is obtained in a straightforward way once $P^{(k)}_{\pm}(c,L)$ are
determined, by imposing the normalization condition  \eqref{qnorm}
order by order in $\lambda^{-1}$.

These results are readily applied, through Eq.\eqref{rf}, to write
down the  large-$\lambda$ expansions of the individual spectral determinants
\begin{eqnarray}\label{logdd}
\partial_\lambda\log D_{\pm}(\lambda) \asymp  L-1+\log (4) +\frac{-1\pm
2}{8\,\lambda} -\pi^2\ \sum_{k=2}^{\infty}P^{(k)}_{\mp}(0,L)\
\lambda^{-k}\,,
\end{eqnarray}
where $L$ is the logarithm \eqref{Ldef},
and the terms $\propto
\lambda^{-2}$ and higher involve the polynomials \eqref{pk}
specified to $c=0$. This equation determines the large-$\lambda$
expansions of $D_{\pm}(\lambda)$ up to an overall numerical
factors
\begin{eqnarray}\label{dassm}
D_{\pm}(\lambda) \asymp d_{\pm}\,
\big(8\pi\re^{-2+\gamma_E}\big)^{\lambda}\
(-\lambda)^{\lambda-\frac{1}{8}\,\pm\, \frac{1}{4}}\ \exp\bigg[\
\sum_{k=1}^{\infty}\,F_{\pm}^{(k)}(L)\,\lambda^{-k}\ \bigg]\,,
\end{eqnarray}
where the polynomials $F_{\pm}^{(k)}(L)$ are easily deducible from
\eqref{logdd}, e.g.
\begin{eqnarray}
F_{\pm}^{(1)}(L) = \mp\,\frac{1}{4\pi^2}\,, \qquad
F_{\pm}^{(2)}(L) = \frac{1 \pm 12\,L}{48\pi^4}\,, \qquad
\text{etc}\ .
\end{eqnarray}
One immediate consequence of the asymptotic expansions\
\eqref{dassm} are analytical  predictions for the regularized sum
\bea\label{lsls} G_+^{(1)}\equiv \sum_{m=0}^{\infty}\,\bigg[\,
\frac{1}{\lambda_{2m}}-\frac{1}{m+1}\,\bigg]\ ,\ \ \ \ \
G_-^{(1)}\equiv \sum_{m=0}^{\infty}\,\bigg[\,
\frac{1}{\lambda_{2m+1}}-\frac{1}{m+1}\,\bigg]\ . \eea
The form of the pre-exponential factor in
\eqref{dassm} implies
\begin{eqnarray}\label{fatey}
G_+^{(1)}= \log(8\pi)-1\,, \qquad G_-^{(1)}= \log(8\pi)-3\, .
\end{eqnarray}

The numerical factors $d_{\pm}$ can not be obtained from
\eqref{logdd}. In fact, at the moment we do not have analytic
expressions for these constants. However, the exact relation
\begin{eqnarray}\label{doverd}
{d_{-}\over d_{+}} ={\sqrt{2}\over \pi}
\end{eqnarray}
follows from\ \eqref{slsssosil}. Note that the constants
$d_{\pm}$ can be written as (fast convergent) products
\begin{eqnarray}\label{dprod}
d_{+} = \frac{\Gamma({3\over 8})}{\sqrt{2\pi}}\,\prod_{m=0}^{\infty}\,
\frac{m+{3\over 8}}{\lambda_{2m}}\ , \qquad
d_{-}=\frac{\Gamma({7\over 8})}{\sqrt{2\pi}}\,\prod_{m=0}^{\infty}\,
\frac{m+{7\over 8}}{\lambda_{2m+1}}\ .
\end{eqnarray}
The relation \eqref{doverd} is a rather nontrivial prediction of
our theory. Fortunately,  the constants $d_{\pm}$ play no role in
derivation of the large-$\lambda$ expansion of the spectrum below.

It is important that the form \eqref{qpmass} was designed to
describe the asymptotic behavior of $Q_{\pm}(\nu|\lambda)$ at
large {\it negative} $\lambda$, therefore \eqref{logdd} generates
the asymptotic expansions of the spectral determinants at
$\lambda\to -\infty$. In view of the analytic structure
\eqref{dpm}, these expansions are in fact valid at all
(sufficiently large) complex $\lambda$, except for when $\lambda$
lies in a narrow sector around the positive real axis in the
complex $\lambda$-plane. But since the main object of our
interest is the spectrum $\{\lambda_n\}$, we are especially
interested in the asymptotics of $D_{\pm}(\lambda)$ at real
positive $\lambda$. Below we argue that the asymptotic behavior
in this domain is correctly described as
\begin{eqnarray}\label{dassp}
D_{\pm}(\lambda) \asymp D_{\pm}^{(+)}(\lambda) +
D_{\pm}^{(-)}(\lambda)\,,
\end{eqnarray}
where $D_{\pm}^{(+)}(\lambda)$ and $D_{\pm}^{(-)}(\lambda)$ are
the results of term-by-term analytic continuations of the series
\eqref{dassm} from the negative to the positive part of the real
axis, in the clockwise and the counterclockwise directions,
respectively (formally, $D_{\pm}^{(+)}(\lambda) =
D_{\pm}(-\re^{-\ri\pi}\lambda)$ and $D_{\pm}^{(-)}(\lambda) =
D_{\pm}(-\re^{\ri\pi}\lambda)$, where $D_{\pm}(\lambda)$ are
understood as the series \eqref{dassm}). Then we can write the
$\lambda\to +\infty$ expansions as
\begin{eqnarray}
\label{hssatsat}
D_{\pm}(\lambda) \asymp 2\, d_{\pm}\,
\big(8\pi\re^{-2+\gamma_E}\big)^{\lambda}\
 \lambda^{\lambda-\frac{1}{8}\,\pm\,
\frac{1}{4}}\ \
\re^{\Xi_{\pm}(\lambda)}\
\cos\Big[\,{\textstyle \frac{\pi}{2}}\,
\big(\,2\lambda-
{\textstyle\frac{1}{4}}\pm{\textstyle\frac{1}{2}}
-\Phi_{\pm}(\lambda)\,\big)\, \Big]\,,
\end{eqnarray}
where $\Xi_{\pm}(\lambda)$ and $\Phi_{\pm}(\lambda)$ are the
asymptotic series of the form
\begin{eqnarray}
\Xi_{\pm}(\lambda) =\sum_{k=1}^{\infty}\Xi_{\pm}^{(k)}(l)\ \lambda^{-k}\,,
\qquad
\Phi_{\pm}(\lambda)=\sum_{k=2}^{\infty}\Phi_{\pm}^{(k)}(l)\ \lambda^{-k}\, .
\end{eqnarray}
Here the coefficients $\Xi_{\pm}^{(k)}(l)$ and
$\Phi_{\pm}^{(k)}(l)$ are polynomials in the real logarithm
\begin{eqnarray}\label{jaay}
l =\log(2\pi\lambda)+\gamma_E\,,
\end{eqnarray}
directly related to the polynomials $F_{\pm}^{(k)}(L)$ in
\eqref{dassm},
\begin{eqnarray}
\Xi_{\pm}^{(k)}(l)&=&{{1\over 2}}\ \big[\,
F_{\pm}^{(k)}(l+\ri\pi)+F_{\pm}^{(k)}(l-\ri\pi)\, \big]\,,\\
\Phi_{\pm}^{(k)}(l)&=& {{\ri \over \pi}}\ \big[\,
F_{\pm}^{(k)}(l+\ri\pi)-F_{\pm}^{(k)}(l-\ri\pi)\,\big]\,.\nonumber
\end{eqnarray}
Of these, $\Phi_{\pm}^{(k)}(l)$ are especially important since
they enter the ``quantization conditions''
\begin{eqnarray}\label{wkbeven}
&&2\lambda - {\textstyle\frac{3}{4}} - \sum_{k=2}^{\infty}\,\Phi_{+}^{(k)}(l)\
\lambda^{-k} = 2\,m\,,\\
\label{wkbodd}&&2\lambda - {\textstyle\frac{3}{4}} -
\sum_{k=2}^{\infty}\,\Phi_{-}^{(k)}(l)\ \lambda^{-k} = 2\,m+1\,,
\end{eqnarray}
which, with $m=0,\,1,\,2\ldots$\ , determine the eigenvalues
$\lambda_{2m}$ and $\lambda_{2m+1}$, respectively. Therefore we
present explicitly $\Phi_{\pm}^{(k)}(l)$ up to $k=7$ in Appendix
A (see Eqs.\eqref{lssls} and \eqref{slsuays}).
The large-$n$ expansion \eqref{semiodd} follows directly from
\eqref{wkbeven},\,\eqref{wkbodd}.

At the moment we do not have completely satisfactory proof of
\eqref{dassp}. However there is a body of supporting arguments.
The most important concerns the behavior of the functions
$Q_{\pm}(\nu|\lambda)$ themselves at large positive $\lambda$.
The easiest way to understand the situation is again through the
analytic continuation in $\lambda$. The expression \eqref{qpmass}
can be analytically continued to positive $\lambda$ term by term
in the expansions \eqref{uhat} and \eqref{ksaksas}. With any such
continuation, it still satisfies the functional equation
\eqref{fe1} order by order in $\lambda$, and its coefficients are
still free of poles in the strip $|\Im m\,\nu|\leq 2$. But there
are two natural ways of the continuation - one is through the
upper half-plane, and another is through the lower one. Thus at
positive $\lambda$ we have two series-like solutions of
\eqref{fe1}, with correct analyticity in the strip $|\Im
m\,\nu|\leq 2$, both for $Q_{+}$ and $Q_{-}$. Let us denote them
$Q_{\pm}^{(+)}(\nu|\lambda)$ and $Q_{\pm}^{(-)}(\nu|\lambda)$.
The problem is that each of them exhibits unacceptably rapid
growth at $|\Re e\,\nu|\gg 1$. It is possible to show that at,
say, positive $\nu \to  +\infty$ they behave as
\begin{eqnarray}\label{growexp}
Q_{\pm}^{(+)}(\nu|\lambda) &\to&
\re^{\ri\pi (\lambda+{1\over 8})}\   R_\pm (-2\pi\lambda,\lambda)|_{L=l-\ri\pi}\
\ \ \ \re^{\pi\nu}\ {\tilde M} (\nu,2\pi\lambda)\, ,\nonumber\\
Q_{\pm}^{(-)}(\nu|\lambda) &\to& \pm   \re^{-\ri\pi(\lambda+{1\over 8})}\
R_\pm (2\pi\lambda, \lambda)|_{L=l+\ri\pi}\ \ \
\re^{\pi\nu}\ {\tilde M} (\nu,2\pi\lambda)\,.
\end{eqnarray}
Here $ {\tilde M} $ is a certain combination of the hypergeometric
functions $M_\pm(\nu,2\pi\lambda)$ \eqref{up} with coefficients
which, unlike exponential factors $\re^{\pm \ri\pi\lambda}$, admit
large-$\lambda$ expansion similar to\ \eqref{rexp}. Note that this
behavior is completely compatible with the functional equation,
but contradicts the required large-$\nu$ behavior of true
functions $Q_{\pm}(\nu|\lambda)$, which must grow slower then any
exponential. Admittedly, we are dealing here with asymptotic
series in $\lambda^{-1}$, and using them to judge the
$\nu\to\infty$ asymptotics is problematic. But the series in
\eqref{uhat} is expected to be approximative at large $\lambda$ as
long as $\sqrt{\lambda}\gg \nu$ (and even more so for the series
\eqref{rexp}). If one focuses on the region $\sqrt{\lambda}\gg \nu
\gg 1$, the exponential growth \eqref{growexp} is clearly
incompatible with the expected behavior of true functions
$Q_{\pm}(\nu|\lambda)$, which at $\nu\gg 1$ must quickly (with
exponential accuracy) become linear combinations of the functions
$M_{+}(\nu, 2\pi\lambda)$ and $M_{-}(\nu, 2\pi\lambda)$ defined
in \eqref{up}. However, it is clear from \eqref{growexp} that one
can form special linear combinations of $Q^{(+)}$ and $Q^{(-)}$
in which the growing terms cancel out. These combinations involve
the factors $\re^{\ri\pi\lambda}$ and $\re^{-\ri\pi\lambda}$
which do not admit $\lambda^{-1}$ expansions; this is why
straightforward $\lambda^{-1}$ expansions are impossible at
positive $\lambda$. The most compact way to describe these linear
combinations is in terms of somewhat differently normalized
functions
\begin{eqnarray}\label{calq}\label{qtdef}
{\cal Q}_{\pm}(\nu|\lambda) =2^{\pm 1}\ \ri^{\pm 1-1\over 2}\
D_{\pm}(\lambda)\ Q_{\pm}(\nu|\lambda)\,;
\end{eqnarray}
instead of \eqref{qnorm} they satisfy
\begin{eqnarray}\label{qtnorm}
{\cal Q}_{+}(2\ri|\lambda) =2\  D_{+}(\lambda)\,, \qquad {\cal
Q}_{-}(0|\lambda) ={{1\over 2\ri}}\  D_{-}(\lambda)\,.
\end{eqnarray}
The normalization factors in \eqref{calq} make ${\cal
Q}_{\pm}(\nu|\lambda)$ entire functions of $\lambda$. In
particular, instead of \eqref{lambdapoles}, at the spectral values
of $\lambda$ we simply have
\begin{eqnarray}
\Psi_{2m}(\nu)\, \propto\,  \frac{{\cal
Q}_{+}(\nu|\lambda_{2m})}{\nu\,\cosh(\frac{\pi\nu}{2})}\,, \qquad
\Psi_{2m+1}(\nu)\, \propto\,  \frac{{\cal
Q}_{-}(\nu|\lambda_{2m+1})}{\nu\,\cosh(\frac{\pi\nu}{2})}\ .
\end{eqnarray}
At negative $\lambda$ (indeed, at all complex $\lambda$ except
for the narrow sector around the positive real axis), the
large-$\lambda$ expansion can still be written in the form
\eqref{qpmass}, with the coefficients $R_{\pm}(z,\lambda)$
replaced by
\begin{eqnarray}\label{calr}
{\cal R}_{\pm}(z,\lambda)= d_\mp\,
\big(8\pi\re^{-2+\gamma_E}\big)^{\lambda}\
(-\lambda)^{\lambda-\frac{5}{8}\,\mp\, \frac{1}{4}}\ \bigg[\,1 +
\sum_{k=1}^{\infty}\,{\cal R} _{\pm}^{(k)}(c, L)\,\lambda^{-k}\,
\bigg]\,,
\end{eqnarray}
with new polynomials ${\cal R}_{\pm}(c,L)$ which are obtained by
combining \eqref{ksaksas} with \eqref{dassm}. Now, let ${\cal
Q}_{\pm}^{(+)}(\nu|\lambda)$ and ${\cal
Q}_{\pm}^{(-)}(\nu|\lambda)$ be two asymptotic expansions
obtained by formal term-by-term analytic continuation in
$\lambda$ from negative to positive $\lambda$, one through the
upper half-plane and another through the lower one. It turns out
that it is exactly the sums ${\cal Q}_{\pm}^{(+)} + {\cal
Q}_{\pm}^{(-)}$ in which the unacceptable growing terms
\eqref{growexp} cancel out. Thus it is natural to assume that
correct asymptotic behavior of true functions ${\cal
Q}_{\pm}(\nu|\lambda)$ at real positive $\lambda$ is given by
these sums \footnote{The situation is reminiscent to how the WKB
expansions of the wave-functions in quantum mechanics are matched
around the turning points.},
\begin{eqnarray}\label{qpmassp}
{\cal Q}_{\pm}(\nu|\lambda) \asymp {\cal
Q}_{\pm}^{(+)}(\nu|\lambda)+{\cal Q}_{\pm}^{(-)}(\nu|\lambda)\,,
\qquad \lambda\to +\infty\,.
\end{eqnarray}
From this, the form \eqref{dassp} immediately follows. We note
here that after the cancellation of the growing terms, the
combinations \eqref{qpmassp} have the following behavior at real
$|\nu|\gg 1$,
\begin{eqnarray}
{\cal Q}_{+}(\nu,\lambda) \sim   M_+(|\nu|, 2\pi\lambda)\ ,\ \ \ \ \
{\cal Q}_{-}(\nu,\lambda) \sim  \sgn(\nu)\  M_+(|\nu|, 2\pi\lambda)\ .
\end{eqnarray}
It turns out that these equations give very good approximations of
the functions even at $\nu \sim 1$, and even if $\lambda$ is not
particularly large (see next section).

\begin{figure}[!ht]
\centering
\includegraphics[width=9  cm]{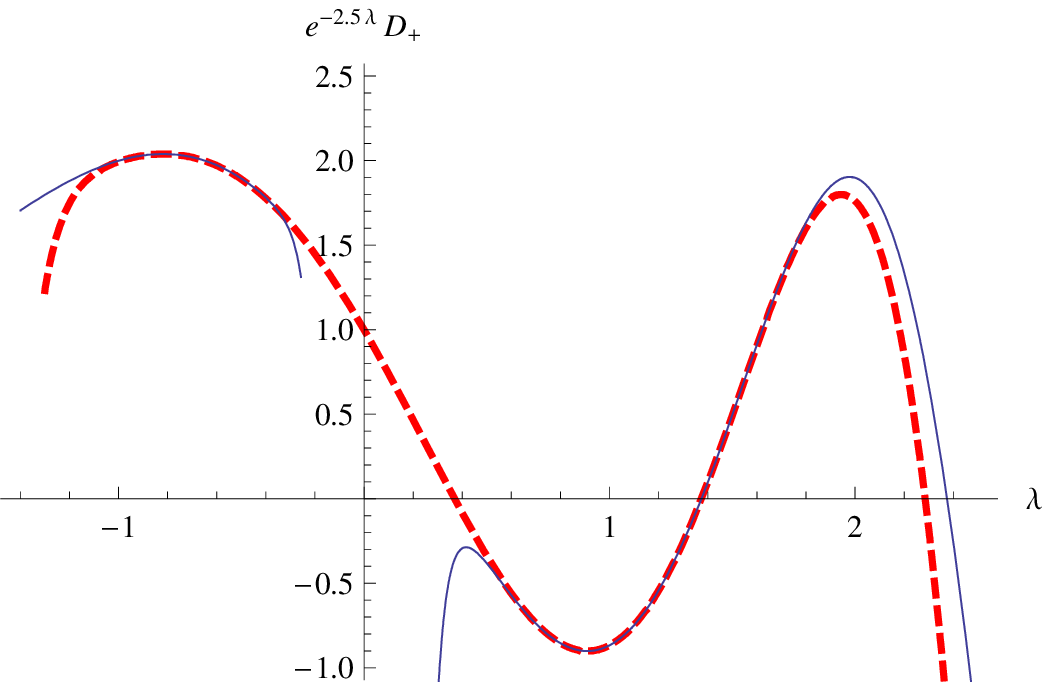}
\caption{
Plots of small- and large-$\lambda$ expansions of
$\re^{-2.5\lambda}\,D_+$. The $\lambda$-expansion, with terms up
to $\propto \lambda^{14}$ in \eqref{dpmser}, is shown as the
dashed line. The solid lines represent the large-$\lambda$
expansions, i.e. \eqref{dassm} at negative $\lambda$, and
\eqref{hssatsat} at positive $\lambda$; in both cases terms up to
$\propto \lambda^{-6}$ are included.}
\label{fig1}
\includegraphics[width=9cm]{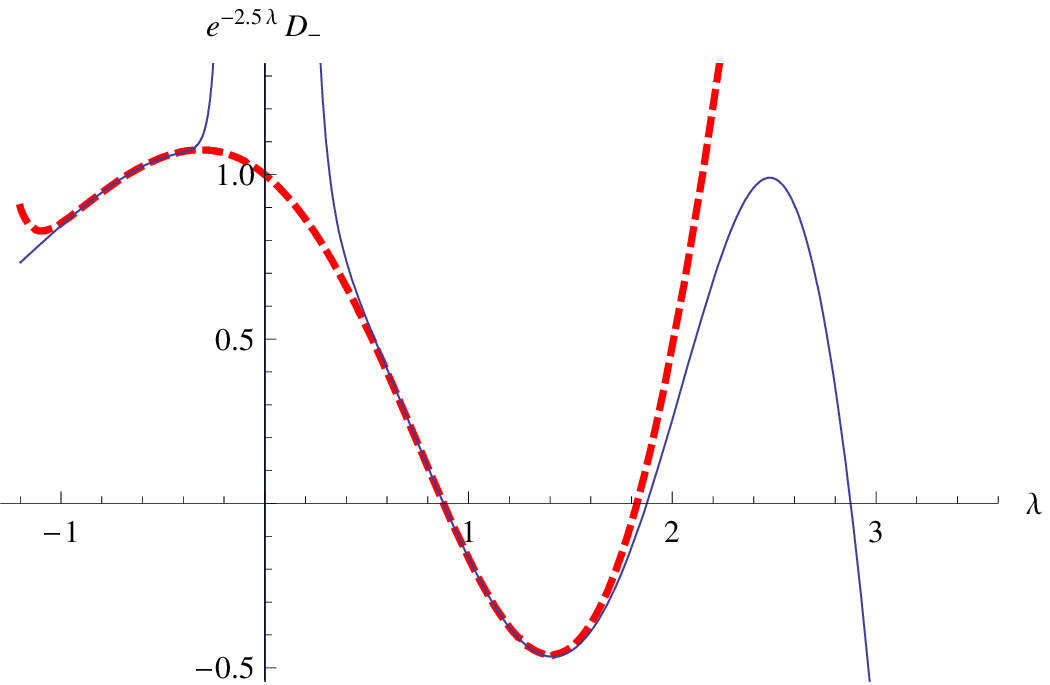}
\caption{
Same as in Fig.\,1, but for
$\re^{-2.5\lambda}\,D_-$. In this case the small-$\lambda$
expansions is truncated to terms $\propto \lambda^{13}$. }
\label{fig2}
\end{figure}
Another piece of evidence supporting \eqref{dassp} is numerical.
First of all, the numerical values of $\lambda_{2m}$ and
$\lambda_{2m+1}$ obtained from  \eqref{wkbeven} and
\eqref{wkbodd}, with some reasonable number of terms in the
$\lambda^{-1}$ expansions included, provide remarkably accurate
estimates for the eigenvalues, even for low levels. We discuss
these numerics in the next section. But one can also match the
large-$\lambda$ expansions of the spectral determinants to the
power series expansions
\begin{eqnarray}\label{dpmser}
D_{\pm}(\lambda) =1+(1\mp 1)\lambda+
\sum_{s=2}^{\infty}\,D_{\pm}^{(s)}\ \lambda^s\,.
\end{eqnarray}
The latter converge in the whole $\lambda$ plane. With many terms
included, the expansions \eqref{dpmser} are expected to
approximate the functions $D_{\pm}(\lambda)$ well even if
$|\lambda|$ is not small. Since we know as many as 13 terms of
\eqref{dpmser}, one expects to have substantial domain at
negative $\lambda$ where the truncated series \eqref{dpmser}
match the asymptotic expansions \eqref{dassm} (again, with a
reasonable number of terms in the sum). More crucially, if
\eqref{dassp} is correct, there must be substantial domain of
positive $\lambda$ where it matches \eqref{dassp}. This comparison
requires knowing the constants $d_{\pm}$ in
Eqs.\eqref{dassm},\,\eqref{dassp}. We use here the numerical
estimates from the next section (see Eq.\,\eqref{sluyyssa}). In
Figs.\ref{fig1},\,\ref{fig2} we present simultaneous plots of the
$\lambda$ expansions \eqref{dpmser}, with as many terms as are
available, and the large-$\lambda$ expansions \eqref{dassm} and
\eqref{hssatsat}, with the sums including all terms up to
$\propto\lambda^{-6}$. In fact, the plots are for
$\re^{-2.5\,\lambda}\,D_{\pm}(\lambda)$, with the exponential
factor added to make interesting parts of all three plots visible
in the same picture ($D_{\pm}(\lambda)$ themselves develop  large
amplitudes already at $\lambda\sim 1$). As expected, there is a
good match at negative $\lambda$ between $-1$ and $-0.4$, but one
can also see a clear match at positive $\lambda$, in the domain
between $0.6$ and $1.6$ where the functions already show ``live''
behavior. Note that the two lowest zeros of both $D_{+}$ and
$D_{-}$ are already visible at these orders of the
$\lambda$-expansions. In fact, the positions of the lowest zeros
$\lambda_0$ and $\lambda_1$ stabilize rather fast as one adds
more and more terms to \eqref{dpmser}. This convergence is
particularly impressive for $\lambda_0$. Pad${\acute {\rm e}}$
approximation  of the $\lambda$-expansion of  the ratio
${D_{+}(\lambda)\over D_{-}(\lambda)}$, Eq.\eqref{qdrel}, yields
the following estimate of the lowest eigenvalue,
\begin{eqnarray}
2\lambda_0 =0.737061746292690\  .
\end{eqnarray}
Compare this number to the numerical result in Table\,1.

\section{Numerical results}

As was mentioned, it is not difficult to compute eigenvalues
$\{\lambda_n\}$ by  direct numerical solution of  equation
\eqref{bs0}. A variety of numerical methods exists
\cite{thooft,Hanson,Huang,Jaffe,Staudacher}. We have used the
expansion of $\varphi(x)$ in Chebyshev polynomials from
\cite{Hanson, Jaffe}, which seems particularly suitable in the
case \eqref{alpha0} since it automatically guarantees the function
$\varphi(x)$ correct behavior near the boundaries $x=0,1$
(besides, its implementation requires perhaps the least amount of
creative programming). With this method, fourteen significant
digits for as many as 50 lowest eigenvalues $\lambda_n$ can be
obtained by truncating to matrices of the size $400\times 400$.
Below we use the notation $\lambda_{n}^{\text{(num)}}$ for these
numerical estimates.

\begin{table}[!ht]
\begin{center}
\begin{tabular}{| c | l | l | l || l|}
\hline \rule{0mm}{3.6mm}
$n$ & $2\lambda_{n}$ from Eq.\,\eqref{semiodd}
 & $2\lambda_{n}^{(7)}$ & $2\, \delta\lambda_{n}^{(7)}$ &  $2\lambda_{n}^{(\text{num})}$\\
\hline
$0$  & 0.730                & ******                    &  ******               & 0.73706174629269 \\
$2$  & 2.748145              & 2.748159         &$6.3\times 10^{-5}$  & 2.7481609123706 \\
$4$  & 4.749299             & 4.7492955        &$1.8\times 10^{-6}$  & 4.7492953810375 \\
$6$  & 6.749631             & 6.74962943       &$1.7\times 10^{-7}$  & 6.7496294196488 \\
$8$  & 8.7497729             & 8.749771584      &$2.9\times 10^{-8}$  & 8.7497715807892 \\
$10$ & 10.7498458           &10.7498450900     &$6.7\times 10^{-9}$  & 10.749845089160 \\
$12$ & 12.7498885           &12.7498880086     &$2.0\times 10^{-9}$  & 12.749888008416\\
$14$ & 14.7499156           &14.74991524453    &$6.9\times 10^{-10}$ & 14.749915244446\\
$16$ & 16.7499338           &16.74993361109    &$2.7\times 10^{-10}$ & 16.749933611057\\
$18$ & 18.74994673          &18.74994658405    &$1.2\times 10^{-10}$ & 18.749946584034\\
$20$ & 20.74995619          &20.749956088181   &$5.4\times 10^{-11}$ & 20.749956088173\\
$22$ & 22.74996334          &22.749963259765   &$2.7\times 10^{-11}$ & 22.749963259761\\
$24$ & 24.74996886          &24.749968804885   &$1.4\times 10^{-11}$ & 24.749968804883\\
$26$ & 26.74997323          &26.749973181147   &$7.5\times 10^{-12}$ & 26.749973181145\\
$28$ & 28.74997673          &28.749976695732   &$4.2\times 10^{-12}$ & 28.749976695731\\
\hline
\end{tabular}
\end{center}
\caption{ Numerical values of the even eigenvalues $2\lambda_{n}$
from the large-$\lambda$ expansion. The first column gives simply
the numerical values of \eqref{semiodd}, with all higher
corrections ignored. $2\lambda_{n}^{(7)}$ are obtained from
\eqref{wkbeven} with the sum truncated beyond the term $\propto
\lambda^{-7}$. The differences  $2\delta\lambda_{n}^{(7)}=
2\lambda_{n}^{(7)}- 2\lambda_{n}^{(6)}$ are given in the third
column, they show the effect of the term $\propto \lambda^{-7}$.
In the last column we present the eigenvalues
$2\lambda_{n}^{(\text{num})}$ computed by direct numerical
solution of \eqref{bs0}.} \label{leven}
\begin{center}
\begin{tabular}{| c | l | l | l ||l |}
\hline \rule{0mm}{3.6mm}
$n$ & $2\lambda_{n}$ from Eq.\,\eqref{semiodd}
 & $2\lambda_{n}^{(7)}$ &$2\,\delta \lambda_{n}^{(7)}$ &  $2\lambda_{n}^{(\text{num})}$\\
\hline
$1$  & 1.75381    & 1.75396        &$-9.3\times 10^{-4}$ & 1.7537313369175 \\
$3$  & 3.751045   & 3.7510570      &$-8.6\times 10^{-6}$ & 3.7510575817054 \\
$5$  & 5.750487   & 5.75049257     &$-5.0\times 10^{-7}$ & 5.7504926236487 \\
$7$  & 7.7502819  & 7.750284389    &$-6.4\times 10^{-8}$ & 7.7502843971925 \\
$9$  & 9.7501838  & 9.750185133    &$-1.3\times 10^{-8}$ & 9.7501851352539 \\
$11$ & 11.7501294 & 11.7501301421  &$-3.4\times 10^{-9}$ & 11.750130142515 \\
$13$ & 13.7500960 & 13.7500965038  &$-1.1\times 10^{-9}$ & 13.750096503972 \\
$15$ & 15.7500741 & 15.75007442838  &$-4.0\times 10^{-10}$ & 15.750074428438 \\
$17$ & 17.7500589 & 17.75005915901   &$-1.6\times 10^{-10}$ & 17.750059159035 \\
$19$ & 19.75004800   & 19.750048157159  &$-7.2\times 10^{-11}$ & 19.750048157169 \\
$21$ & 21.75003985    & 21.750039967124  &$-3.4\times 10^{-11}$ & 21.750039967130 \\
$23$ & 23.75003362   & 23.750033705315  &$-1.7\times 10^{-11}$ & 23.750033705317 \\
$25$ & 25.75002874   & 25.750028810058 &$-9.0\times 10^{-12}$ & 25.750028810060 \\
$27$ & 27.75002486   & 27.750024910393 &$-4.9\times 10^{-12}$ & 27.750024910394  \\
$29$ & 29.75002171   & 29.750021753287 &$-2.7\times 10^{-12}$ & 29.750021753287  \\
\hline
\end{tabular}
\end{center}
\caption{ The same as in Table 1, but for the odd eigenvalues $2\lambda_n$.}
\label{lodd}
\end{table}

In Tables 1 and 2 we compare these numbers, for even and odd $n$
separately, with the results of large-$\lambda$ expansions. The
first column in each of these  tables shows numerical values
yielded by Eq.\eqref{semiodd}, with all terms explicitly written
there included. One can observe significant improvement as
compared to the leading semiclassical approximation $\lambda_n
\approx n+{3\over 4}$, even for the low levels such as
$\lambda_1$ and $\lambda_2$. The approximation \eqref{semiodd}
corresponds to truncating the sums in Eq.\eqref{wkbeven} and
\eqref{wkbodd} to terms $\sim \lambda^{-3}$, but one can obtain
further corrections by including higher-order terms.  We denote
$\lambda_{n}^{(k)}$ the estimates from  equations
\eqref{wkbeven}, \eqref{wkbodd} with terms up to
$\propto\lambda^{-k}$ included, and  present  the numerical
values of $\lambda_{n}^{(7)}$ (together with the deviations
$\delta\lambda^{(7)}_n= \lambda_{n}^{(7)}-\lambda_{n}^{(6)}$ to
show the expected accuracy of this approximation). Since we are
dealing with an asymptotic expansion, one does not expect it to
work well for low levels, but Tables 1,\,2 show that including
these further corrections results in noticeable improvement even
for levels as low as $\lambda_3$ and $\lambda_4$, and for higher
levels the improvement becomes impressive. For $n \geq 30$
$\lambda_{n}^{(7)}$ are indistinguishable from $\lambda_{n}^{\rm
(num)}$ within the accuracy of the latter.

Another impressive agreement is in terms of the sum rules
\eqref{gpm},\,\eqref{lsls}. One can evaluate the spectral sums in
\eqref{gpm},\,\eqref{lsls} using the numerical values
$\lambda_{n}^{(\text{num})}$, and compare these numerical
estimates $\big[\,G_{\pm}^{(s)}\,\big]^{(\rm num)}$ with the
analytic predictions \eqref{gpmfew},\,\eqref{fatey} and
\eqref{ssas},\,\eqref{ssasusy}. In fact, for low $s$ the sums do
not converge that fast. For instance, to estimate
$\big[\,G_{\pm}^{(2)}\,\big]^{(\rm num)}$ to fourteen digits one
needs to include as many as $10^7$ eigenvalues. Of course, this
problem is easy to solve since we have very good large-$n$
asymptotic approximations. In the sums \eqref{gpm}, starting from
some sufficiently large $n$ one simply replaces
$\lambda_{n}^{(\text{num})}$ by the asymptotic form, say
$\lambda_{n}^{(7)}$. In Table 3 we show numerical estimates
obtained in this way for $s=1,2,\ldots,8$. It is an easy and
pleasant exercise to check that these numbers agree with the
analytic expressions \eqref{gpmfew},\,\eqref{fatey} and
\eqref{ssas},\,\eqref{ssasusy} to all digits presented. As was
mentioned, we actually have analytic expressions for
$G_{\pm}^{(s)}$ with $s$ up to 13, and we have verified similar
agreement for these higher values of $s$ as well.
\begin{table}[ht]
\begin{center}
\begin{tabular}{| c | l || l |}
\hline \rule{0mm}{3.6mm}
$s$ & $\big[\,G_{+}^{(s)}\,\big]^{(\rm num)}$ & $\big[\,G_{-}^{(s)}\,\big]^{(\rm num)}$ \\
\hline
$1$ & 2.22417142752923 & 0.2241714275292 \\
$2$ & 8.4143983221171  & 2.0000000000000 \\
$3$ & 20.4981207536828 & 1.7198241782619  \\
$4$ & 54.538349992708  & 1.7952480377615 \\
$5$ & 147.32680373214  & 1.9789889429098 \\
$6$ & 399.32397653715  & 2.2250507748184 \\
$7$ & 1083.2464075913  & 2.521777906136 \\
$8$ & 2939.1433918727  & 2.867885373267 \\
\hline
\end{tabular}
\end{center}
\caption{Numerical values of of the spectral sums \eqref{gpm},\,\eqref{lsls}.}
\label{F-Cyns}
\end{table}
We also have computed the products \eqref{dprod} with the numerical
spectrum,
\begin{eqnarray}
\label{sluyyssa}
d_{+}^{(\text{num})} =0.963178456398\,,\ \  \qquad d_{-}^{(\text{num})}=0.433582639833\, .
\end{eqnarray}
Again, it is easy to check that these numbers
comply with \eqref{doverd} to twelve digits.

\begin{figure}[ht]
\centering
\includegraphics[width=10cm]{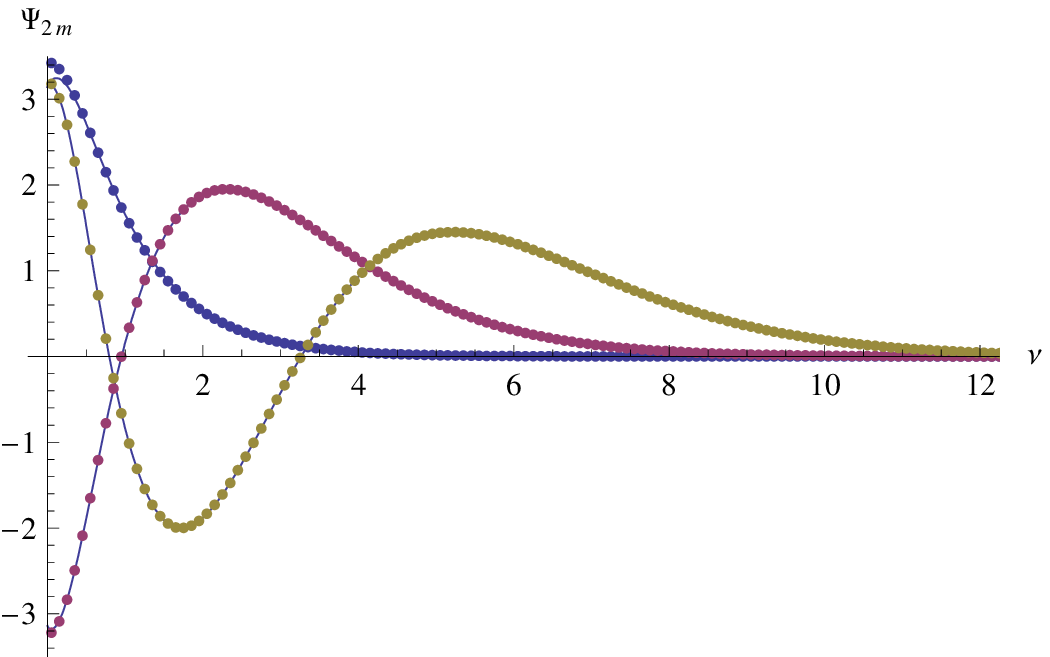}
\caption{Comparison of the approximation \eqref{skklskssa}
for for even eigenfuctions ${ \Psi}_{n}(\nu)$ with $n=0,\,2,\,4$
(solid lines) with the results of direct numerical solution of
\eqref{bsnu} (bullets).}
\label{fig3}
\includegraphics[width=10cm]{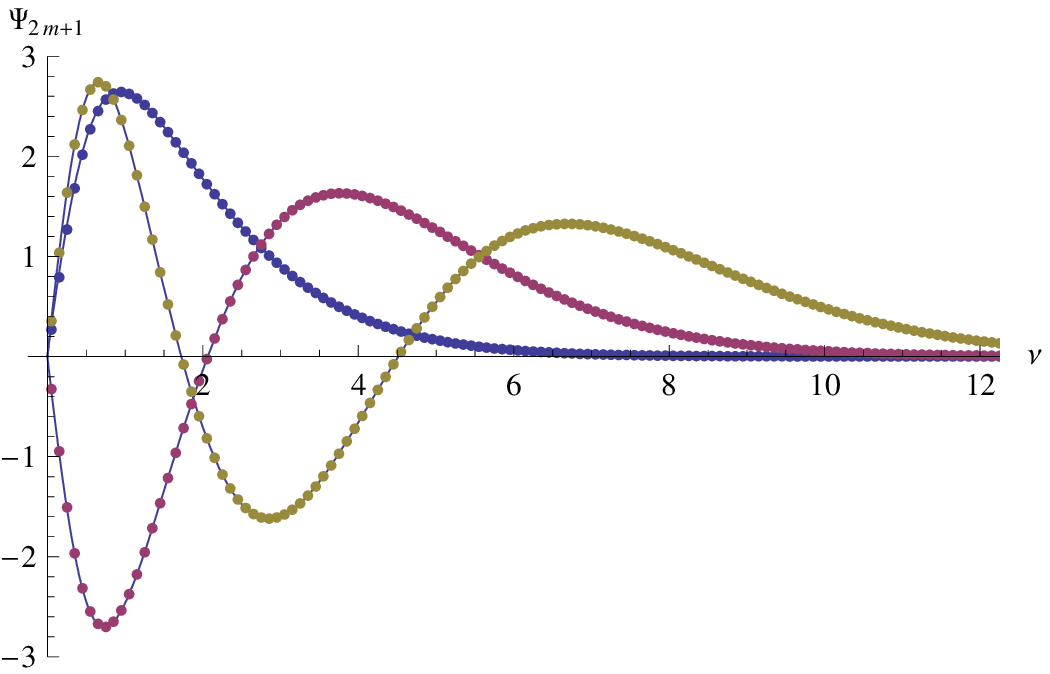}
\caption{
Comparison of the approximation \eqref{skklskssa}
for for odd eigenfuctions ${ \Psi}_{n}(\nu)$ with $n=1,\,3,\,5$
(solid lines) with the results of direct numerical solution of
\eqref{bsnu} (bullets).}
\label{fig4}
\end{figure}

The above numerics concern the eigenvalues $\{\lambda_n\}$. But
it is also interesting to see how the large-$\lambda$ expansions
in Section 4 approximate the associated eigenfunctions
$\Psi_{n}(\nu)$.  It turns out that \eqref{qpmass} provides a
rather good approximation even if one retains only the leading
term 1 in the expansion \eqref{rexp} of the coefficients
$R_{\pm}(z,\lambda)$. In this approximation the sum in
\eqref{uhat} is understood as the hypergeometric function $(-\ri
z)^{1+\frac{\ri\nu}{2}}\,U(1+\frac{\ri\nu}{2},2,-\ri z)$. This
results in the following approximate expression for the normalized
eigenfunctions (which we write for $\nu > 0$; the $\nu < 0$ part
is restored by symmetry),
\begin{eqnarray}\label{skklskssa}
{ \Psi}_{n}(\nu)\approx { \sqrt{8}\, \pi
\lambda_n\ \sinh^2({\pi\nu\over 2})\over \cosh({\pi \nu \over
2})\sqrt{1+\re^{-\pi\nu}}}\ \ \Re e\bigg[\, \ri^{n}\
\re^{-\ri\Phi(\nu)}\ \Gamma\big({\textstyle {i\nu\over 2}}\big)\
U \Big( 1+{\textstyle{\ri\nu\over 2}},\, 2,\,
-2\ri\pi\lambda_n\tanh\big( {\textstyle{\pi\nu\over 2}}\big)
\,\Big)\, \bigg]\, .
\end{eqnarray}
Here the phase $\Phi(\nu)$ has expression
\begin{eqnarray}\label{kslsk}
\Phi(\nu)=\sgn(\nu)\ \bigg[\,{\pi\over 8}-{1\over 4\pi}\
\re^{-\pi|\nu|}\ \Phi\big(\re^{-2\pi|\nu|},\,2,
\,{\textstyle{1\over 2}}\,\big)\ \bigg]\ ,
\end{eqnarray}
in terms of the Lerch transcendent
$\Phi(z,s,a)=\sum_{k=0}^\infty{z^k\over (k+a)^s}$. The
approximation is not expected to be very accurate at small $\nu$,
because \eqref{skklskssa} has a term with singularity
$\sim\nu\,\log\nu$ at $\nu=0$, while true eigenfunctions
$\Psi_n(\nu)$ are analytic at all real $\nu$ (recall that the
higher order terms in \eqref{rexp} were designed precisely to fix
this analytic deficiency). However, numerically the deviations of
\eqref{skklskssa} from true eigenfunctions are rather small even
at small $\nu$. Figs.\,\ref{fig3},\,\ref{fig4} show plots of
\eqref{skklskssa} for low $n$ against the corresponding
eigenfunctions obtained by numerical solution of \eqref{bsnu}.
Deviation at small $\nu$ is barely visible only for $n=0$.

\section{Remarks}

As was mentioned in the introduction, the techniques developed
here extend to the case \eqref{alfalfa}. While we plan to treat
this case in a separate paper, let us announce here some
preliminary results. The large-$\lambda$ expansion generalizes in
an almost straightforward way, yielding the asymptotic large-$n$
expansion of $\lambda_n (\alpha)$. The large-$n$ behavior follows
from the ``quantization condition'', generalizing
\eqref{wkbeven},\,\eqref{wkbodd} in Section 4,
\begin{eqnarray}\label{semialpha}
2\lambda&-&{2\alpha\over\pi^2}\
\log(2\lambda)-C_0(\alpha)+{\alpha^2\over \pi^4\lambda}+ {1\over
2\pi^6\ \lambda^2}\, \Big[\, \alpha^3+ (-1)^n\,\pi^2\,
(1+\alpha)\,\Big]\nonumber \\ &+& {1\over 12\pi^8\,\lambda^3}\,
\Big[\,      5\alpha^4+\pi^2\, (1+\alpha)^2 - (-1)^n\, 12\pi^2\,
(1+\alpha)\, \big(
\log\big(2\pi\re^{\gamma_E}\lambda\big)-C_1(\alpha)\big)\,\Big]\nonumber\\
&+& O\big(\lambda^{-4}\log^2(\lambda)\big)=n\, ,
\end{eqnarray}
where
\begin{eqnarray}\label{c0}
C_0(\alpha) &=&{3\over 4}+ {2\alpha\over\pi^2}\
\log\big(4\pi\re^{\gamma_E}\big)- {\alpha^2\over
2\pi^2}\int_{-\infty}^\infty\rd t\,{\sinh( t)\, (\, \sinh(2 t)-
2 t)\over  t\,\cosh^2(t)\, \big(\,\alpha\,  \sinh(t)+t\, \cosh(t)\,\big)}\, ,\\
\label{c2} C_1(\alpha)&=&\frac{1}{2}+\frac{3\alpha}{2}+
{\alpha\over 8}\ \int_{-\infty}^{\infty}\rd t \ \ {\sinh(2 t)-2
t\over
 t\, \sinh(t)\, \big(\,\alpha\,  \sinh(t)+t\, \cosh(t)\,\big)}\
 .\nonumber
\end{eqnarray}
The first two terms in \eqref{semialpha} are known since
\cite{thooft}. Explicit expression for the constant term $C_0
(\alpha)$, Eq.\eqref{c0}, was previously obtained in
\cite{Brauer} (see also \cite{whothatwas}). We believe the higher
order terms in the expansion in \eqref{semialpha} are new. Further terms
can be derived in a systematic way. Another result
compact enough to be presented here is the analytic expression of
the spectral sums\ \eqref{fatey}\footnote{ This expression
follows in rather straightforward way from analysis in Appendix
B.},
\begin{eqnarray}\label{gpmalpha}
G_{\pm}^{(1)}(\alpha) =
\log(8\pi)-2\pm 1
-{\alpha\over 4}\ \int_{-\infty}^\infty\rd t\
{\sinh(t)\ (\sinh(2 t)\pm 2 t)\over  t\,\cosh^2(t)\, (\alpha\sinh(t)+ t\,\cosh(t)\,)}\ .
\end{eqnarray}

These and other results indicate the rich analytic structure of
$\lambda_n(\alpha)$ as the functions of complex $\alpha$. First,
as expected, $\lambda_n(\alpha)$ have square-root branching point
$\alpha=-1$, which corresponds to the limit $m_1=m_2=0$, where
the chiral symmetry becomes exact. In particular, the lowest
eigenvalue $\lambda_0 (\alpha)$ turns to zero as
$\sqrt{\alpha+1}$. But in addition, there are infinitely many
similar square-root branching points located on the second sheet
of the $\alpha$-plane (i.e. in the left half plane of the
variable $\sqrt{\alpha+1}$), accumulating towards $\alpha=\infty$.
At each of these points one of the even eigenvalues
$\lambda_{2m}(\alpha)$ turns to zero. It is difficult to imagine
that if one takes QCD$_2$ with large but finite $N_c$ these
singularities just disappear. It is more likely that they become
nontrivial critical points of some sort. What are the physics of
these critical points? Can one identify associated (nonunitary)
CFT? These are some of intriguing questions which we plan to
study in the future.

\section*{Acknowledgments}

The authors  would like to thank Alexey Litvinov and Feodor
Smirnov for discussions and interest to this work. VAF
acknowledges kind help of Sergei Meshkov with several numerical
tests. We are grateful to Antonio Pineda for bringing our
attention to important papers \cite{Brauer} and \cite{whothatwas}.

\noindent The research of VAF is supported by the grant
RBRF-CNRS grant PICS-09-02-91064.

\noindent The research of SLL and ABZ is supported in part by DOE grant
$\#$DE-FG02-96 ER 40949.

\section{Appendix A}

{\bf A1}. Analytic expressions for the spectral sums \eqref{gpm}
with $s=2,\,3,\,4$ are given in \eqref{gpmfew}. Here we present
few more expressions for $G_{\pm}^{(s)}$, with $s$ up to 8:
\bea\label{ssas} &&G^{(5)}_+=\frac{2}{135} \big[\, 16 \pi ^4-50
\pi ^2 (2+21 \zeta (3))+105
   \left(20 \zeta (3)+105 \zeta^2 (3)+31 \zeta (5)\right)\, \big]\nonumber \\
&&G^{(6)}_+=\frac{1}{4320} \, \big[\, \pi ^4 (4090+1449 \zeta
(3))-20 \pi ^2 \left(140+9912
   \zeta (3)+2646 \zeta (3)^2+837 \zeta (5)\right)\nonumber \\
&&+ 15
   \left(138768 \zeta^2 (3)+24696 \zeta^3 (3)+16120 \zeta (5)+28
   \zeta (3) (140+837 \zeta (5))+
3429 \zeta (7)\right)\, \big]\nonumber \\
&&G^{(7)}_+=\frac{1}{567000}\, \big[\, -12288 \pi ^6+49 \pi ^4
(21002+45969 \zeta (3))-2940
  \pi^2 \big(36+15400 \zeta (3)\nonumber \\
&&+22050 \zeta^2(3)+4495 \zeta
   (5)\big)
+315 \big(1509200 \zeta^2(3)+1440600 \zeta^3(3)
+60760 \zeta (5)\nonumber\\
&&+196 \zeta (3) (36+4495 \zeta (5))+51689
   \zeta (7)\big)\,\big]
\eea
\bea
&&G^{(8)}_+=\frac{1}{1944000}\, \big[\, -4 \pi ^6 (97286+12375
\zeta (3))+\pi ^4
   \big(4210624+38551128 \zeta (3)\nonumber\\
&&+5622750 \zeta^2(3)
+767250
   \zeta (5)\big)-30 \pi ^2 \big(2464+32104800 \zeta^2(3)+
3704400 \zeta^3 (3)\nonumber \\
&&+4032480 \zeta (5)
+8400 \zeta (3)
   (712+279 \zeta (5))+142875 \zeta (7)\big)+315
   \big(21403200 \zeta^3 (3)\nonumber\\
&&+1852200 \zeta^4 (3)
+76880 \zeta
   (5)+432450 \zeta^2 (5)
+8400 \zeta^2 (3) (712+279 \zeta
   (5))\nonumber\\ &&
+167132 \zeta (7)
+\zeta (3) (4928+8064960 \zeta
   (5)+285750 \zeta (7))+27375 \zeta (9)\big)\, \big]
\nonumber
\eea
\bea\label{ssasusy} &&G^{(5)}_-=\frac{1}{225} \big[\,-32 \pi ^4+50
\pi ^2 (6+7 \zeta (3))-15
   (76+155 \zeta (5))\,\big]
\nonumber\\
&&G^{(6)}_-=\frac{1}{97200}\, \big[\, 681120-214800 \pi ^2+14426
\pi ^4+201600 \pi ^2 \zeta
   (3)-21735 \pi ^4 \zeta (3)\nonumber\\
&&-1339200 \zeta (5)
+251100 \pi ^2
   \zeta (5)-771525 \zeta (7)\, \big]
\nonumber\\
&&G^{(7)}_-=\frac{1}{1190700}\, \big[\, -11541600+4245360 \pi^2
-519302 \pi^4+18432 \pi ^6 \nonumber\\
&&+21609
   \pi ^4 \zeta (3)
+1367100 \pi ^2 \zeta (5)-10921365 \zeta
   (7)\, \big]\\
&&G^{(8)}_-= \frac{1}{76204800}\, \big[\, 1021799520-429522240 \pi
^2
+60393480 \pi ^4-2819800 \pi^6\nonumber\\
&&+110308800 \pi ^2 \zeta (3)
-34223168 \pi ^4 \zeta
   (3)+1455300 \pi ^6 \zeta (3)+25930800 \pi ^4 \zeta^2(3)\nonumber\\
&&-732765600 \zeta (5)
+346332000 \pi^2 \zeta (5)
-22557150
   \pi ^4 \zeta (5)-344509200 \pi^2 \zeta(3) \zeta
   (5)\nonumber\\
&&+1144262700 \zeta^2 (5)
-860267520 \zeta (7)+126015750 \pi
   ^2 \zeta (7)-253519875 \zeta (9)\, \big]\nonumber
\eea

\noindent Expressions for $G_{\pm}^{(s)}$ with even higher $s$
(we have them all the way up to $s=13$) have similar structure,
but appear too cumbersome to fit in reasonable page space.

\bigskip
{\bf A2}. Coefficients $\Phi_\pm^{(k)}(l)$ in
Eqs.\eqref{wkbeven},\,\eqref{wkbodd} for $k\leq 7$:
\bea\label{lssls}
\Phi_+^{(2)}(l)&=&-\frac{1}{2 \pi ^4}\nonumber \\
\Phi_+^{(3)}(l)&=&\frac{12\, l-7}{12 \pi ^6}\nonumber\\
\Phi_+^{(4)}(l)&=&
\frac{1}{16 \pi ^{8}}\ \Big[\,
16 \pi ^2-5+44\, l-24\,l^2\, \Big]\nonumber\\
\Phi_+^{(5)}(l)&=&\frac{1}{12 \pi^{10}}\  \Big[\, 3+76 \pi ^2+12 \zeta(3)+
(48-60\pi^2)\,l -84\, l^2+24\, l^3\,  \Big]\nonumber\\
\Phi_+^{(6)}(l)&=&
\frac{1}{144 \pi^{12}}\ \Big[\,
111+2965 \pi^2-828\pi^4+948 \zeta(3)
+\big(\, 396 - 6228 \pi^2 - 720 \zeta(3)\,\big)\, l\nonumber\\
&+&
(2160\pi^2-2448)\,l^2+1968\, l^3 - 360\, l^4\, \Big]\\
\Phi_+^{(7)}(l)&=&\frac{1}{960 \pi ^{14}}\ \Big[\,
870 + 40865 \pi^2 - 79264 \pi^4 + 21560 \zeta(3)
-
 16800 \pi^2 \zeta(3) + 4320 \zeta(5)\nonumber\\
&-&\big(1800 + 181680 \pi^2 - 47040 \pi^4 + 42720 \zeta(3) \big)\, l+
\big( 160320 \pi^2-24240\nonumber\\
& +& 14400 \zeta(3)\, \big)\, l^2
+
(45760 - 33600 \pi^2)\, l^3 - 22080\, l^4 + 2880\, l^5\, \Big]\ . \nonumber
\eea
\bea\label{slsuays}
\Phi_-^{(2)}(l)&=&\frac{1}{2 \pi ^4}\nonumber \\
\Phi_-^{(3)}(l)&=&\frac{5-12\, l}{12 \pi ^6}\nonumber\\
\Phi_-^{(4)}(l)&=&
\frac{1}{16 \pi ^{8}}\ \Big[\,
3 - 16 \pi^2-36\, l + 24\, l^2\, \Big]\nonumber\\
\Phi_-^{(5)}(l)&=&\frac{1}{12 \pi^{10}}\  \Big[\, -3 - 70 \pi^2 - 12 \zeta(3)+
( 60 \pi^2-36)\, l + 72\, l^2 - 24\, l^3\, \Big]\nonumber\\
\Phi_-^{(6)}(l)&=&
\frac{1}{144 \pi^{12}}\ \Big[\,
828 \pi^4-87 - 2633 \pi^2  - 900 \zeta(3)+
\big(\,5820\pi^2-252  + 720 \zeta(3)\,\big)\, l\nonumber\\
&+&(1944 - 2160 \pi^2)\, l^2-1728\, l^3+360\, l^4\, \Big]\\
\Phi_-^{(7)}(l)&=&\frac{1}{960 \pi ^{14}}\ \Big[\,
76864 \pi^4 -630 - 35355 \pi^2   - 19800 \zeta(3) +
 16800 \pi^2 \zeta(3) - 4320 \zeta(5)\nonumber\\
&+&
\big(1800 + 163920 \pi^2 - 47040 \pi^4 + 40800 \zeta(3)\, \big)\, l+
\big(\,18000 - 151200 \pi^2\nonumber\\
& -& 14400 \zeta(3)\, \big)\, l^2
+
( 33600 \pi^2-37440)\, l^3 + 19680\, l^4 - 2880\, l^5 \, \Big]\ .\nonumber
\eea

\section{Appendix B}

Here we describe some technical details of our analysis of the
integral equation \eqref{unh0} with the r.h.s. \eqref{fpm}. To
make the equations shorter, throughout this appendix we trade the
variable $\nu$ for
\begin{eqnarray}\label{ksssa}
t\equiv{\pi\nu\over 2}\ ,
\end{eqnarray}
but, with some abuse of notations, retain the same symbols for
basic functions. Thus $\Psi_\pm(t|\lambda)$ will stand for
solutions of the integral equations
\begin{eqnarray}\label{lisuyask}
f(t)\, \Psi_+(t|\lambda)-{t\over \sinh(t)} &=& \lambda\
\int_{-\infty}^{\infty}\rd t'\
S(t-t')\ \Psi_+(t'|\lambda)\\
f(t)\, \Psi_-(t|\lambda)-{\pi\over 2\sinh(t)} &=& \lambda\
\dashint_{-\infty}^{\infty}\rd t'\ S(t-t')\ \Psi_-(t'|\lambda)\
,\nonumber
\end{eqnarray}
with the kernel
\begin{eqnarray}\label{lksls}
S(t)={t\over\sinh(t)}\ .
\end{eqnarray}
The analysis below does not depend on a specific form of the
function $f(t)$. With $f(t) = t\,\coth (t)$, Eqs.\eqref{lisuyask}
are equivalent to \eqref{unh0}, \eqref{fpm}, but almost all
statements below remain valid if one takes the more general form
\begin{eqnarray}\label{sssos}
f(t)=\alpha+t\, \coth(t)\ ,
\end{eqnarray}
which appears in analysis of \eqref{bs0} with nonzero but equal
$\alpha_1=\alpha_2 =\alpha$.

Eq.\eqref{lisuyask} defines the spectral problem
\begin{eqnarray}\label{kssksksa}
{\hat K}\, \phi(t)=\lambda^{-1}\ \phi(t)
\end{eqnarray}
for the  Fredholm
operator
\bea\label{lsaoai} {\hat K}\phi(t)\equiv
\int_{-\infty}^{\infty}\rd t'\, K(t,t')\ \phi(t')
\eea
with the kernel
\bea\label{kksa}
K(t,t')={S(t-t')\over \sqrt{f(t)f(t')}}\ ,
\eea
where $\phi=\sqrt{f}\  \Psi$. Let $R(t,t'|\lambda)$ be the
corresponding resolvent, i.e. the kernel of the operator ${{\hat K}\over
1-\lambda {\hat K}}$. By definition, it satisfies the equation
\bea\label{klskks} R(t,t'|\lambda)-\lambda\
\int_{-\infty}^{\infty}\rd \tau\ K(t,\tau)\  R(\tau, t'|\lambda)=
 K(t,t')\ .
\eea
The spectral sums \eqref{gpm} and \eqref{lsls} are related
to the resolvent by the trace identities
\bea\label{lsskla}
\sum_{s=1}^\infty \big[\, G_+^{(s)}+G_-^{(s)}\, \big]\
\lambda^{s-1}&=& {\rm C}+\int_{-\infty}^{\infty} \rd t
\ \big[\, R(t,t|\lambda)-R^{(0)}(t)\,\big]\\
\sum_{s=1}^\infty \big[\, G_+^{(s)}-G_-^{(s)}\, \big]\ \lambda^{s-1}&=
&\int_{-\infty}^{\infty}\rd t \  R(t,-t|\lambda)\ . \label{alasksa}
\eea
The constant C in \eqref{lsskla} depends on the choice of the
subtraction term $R^{(0)}(t)$ needed to make the integral convergent. We take
\bea\label{ksasa} R^{
(0)}(t)=
\frac{\tanh(t)}{t}\ .
\eea
With this choice the constant
can be shown to be exactly
\begin{eqnarray}
{\rm C} = 2\,\log(8\pi)-4\ .
\end{eqnarray}

It is the remarkable property of the kernel \eqref{lksls} in
\eqref{lisuyask} that the resolvent can be expressed in a simple
way through the functions $\Psi_{+}(t|\lambda)$ and
$\Psi_{-}(t|\lambda)$, namely\footnote{In other words, the kernel
\eqref{kksa} belongs to the class of ``integrable'' kernels, see
Ref.\cite{itz} for other kernels with similar property.}
\begin{eqnarray}\label{klslsk}
R(t,t'|\lambda)=
{2\sinh(t)\sinh(t')\over \pi\,\sinh(t'-t)}\
\sqrt{f(t)f(t') }\ \Big[\, \Psi_+ (t'|\lambda)\Psi_-
(t|\lambda)-\Psi_- (t'|\lambda) \Psi_+ (t|\lambda)\, \Big]\ .
\end{eqnarray}

To prove this identity, consider the Liouville - Neumann series
for $\Psi_\pm (t|\lambda)$,
\bea\label{lsaaaiia} f(t)\,
\Psi_+(t|\lambda)&=&\sum_{k=0}^\infty\lambda^k\
\int_{-\infty}^\infty {t_k\over \sinh(t_k) } \
\prod_{j=1}^{k}{\rd t_j\over f(t_j)}\
S(t_{j}-t_{j-1})\\
f(t)\, \Psi_-(t|\lambda) &=& {{\pi\over 2}}\
\sum_{k=0}^\infty\lambda^k\ \dashint_{-\infty}^{\infty}
 {1\over \sinh(t_k) }\prod_
{j=1}^{k} {\rd t_j\over f(t_j)}\ S(t_{j}-t_{j-1})\  ,\nonumber
\eea
where  $t_0\equiv t$. Then we have
\bea\label{kslslsal}
&&{\textstyle{2 \over \pi }}\ f(t)f(t') \Big[\,
\Psi_+({t'}|\lambda)\Psi_-(t|\lambda)- \Psi_-({t'}|\lambda)\Psi_+(
t|\lambda)\, \Big]= \sum_{k
,m}\lambda^{k+m}\dashint_{-\infty}^\infty  \  {\sinh(t'_k-t_m)
\over
\sinh(t'_k)  \sinh(t_m)  }\times\nonumber\\
&&\ \ \ \  S(t'_k-t_m)\  \prod_{j=1}^{k} {\rd t'_j\over f(t'_j)}\
S(t'_{j}-t'_{j-1})\ \ \prod_{i=1}^{m} {\rd t_i\over f(t_i)}\
S(t_{i}-t_{i-1})\ ,
\eea
where again $t_0=t$ and $t_{0}'=t'$.
Let us introduce  uniform notations for the integration variables
\bea\label{slsasa}
(t_1,\ldots t_m;  t'_k,\ldots
t'_1)=(\tau_1,\ldots  \tau_m,\tau_{m+1},\ldots \tau_{k+m}) \ . \eea
The elementary identity \bea\label{ksaks}
\sum_{m=0}^{l}{\sinh(\tau_{m+1}-\tau_m)\over
\sinh(\tau_{m+1})\sinh(\tau_{m})}= {\sinh(\tau_{l+1}-\tau_0)\over
\sin(\tau_{l+1})\sinh(\tau_0)}\ ,
\eea
allows one to put
\eqref{kslslsal} in compact form
\bea\label{kkssasa}
&&{2
\sinh(t)\sinh(t') \over \pi\, \sinh( t'-t) }\
f(t)f( t')\, \big[\, \Psi_+(
t'|\lambda)\Psi_-(t|\lambda)-
\Psi_-( t'|\lambda)\Psi_+(t|\lambda)\, \big]=\nonumber\\
&&\ \ \ \ \ \ \ \ \ \ \ \ \ \ \ \ \ \ \ \ \ \ \ \
\sum_{l=1}^\infty\lambda^l \int_{-\infty}^\infty\prod_{j=1}^{l}
{\rd \tau_j\over f(\tau_j)}\
\prod_{j=1}^{l+1}S(\tau_{j}-\tau_{j-1})\ ,
\eea
where now
$\tau_0\equiv t$, $\tau_{l+1}=t'$. It is easy to see that the
right-hand side here divided by $\sqrt{f(t)f(t')}$ is exactly the Liouville - Neumann series for
the solution of the integral equation\ \eqref{klskks}.

Now, since
\begin{eqnarray}\label{llsasls}
 D_\pm(\lambda) = \Big({8\pi\over\re}\Big)^\lambda\ \exp\bigg[
 -\sum_{s=1}^\infty s^{-1}\ G_\pm^{(s)}\ \lambda^s\, \bigg]\  ,
\end{eqnarray}
combining Eqs.\eqref{lsskla},\,\eqref{alasksa} and \eqref{klslsk} leads
to the following expressions for the logarithmic derivatives of the spectral
determinants,
\begin{eqnarray}\label{kkssa}
&&\partial_\lambda \log(D_+D_-)=2-\qquad\qquad\nonumber\\
&&\ \ \ \int_{-\infty}^{\infty}\rd t\ \bigg\{\, {\pi\over 2 f(t)}\ \
\Big[\, Q_{-}(t|\lambda)\partial_t Q_{+}(t|\lambda)-Q_{+}(t|\lambda)
\partial_t Q_{-}(t|\lambda)\, \Big]-\frac{\tanh(t)}{t}\,\bigg\}
\nonumber\\
\label{kkssb}
&&\partial_\lambda \log\Big({D_+\over D_-}\Big)=
-\int_{-\infty}^{\infty}{\rd t\over f(t)}\ \ \ {\pi\, Q_{+}(t|\lambda)
Q_{-}(t|\lambda) \over \sinh(2t)}\ ,
\end{eqnarray}
where
\begin{eqnarray}\label{slssa}
Q_\pm(t|\lambda)={\textstyle{2\over \pi}}\ \sinh(t)\ f(t)\
\Psi_{\pm}(t|\lambda)\ .
\end{eqnarray}

The above analysis, in particular Eqs.\eqref{kkssb}, applies to
\eqref{lisuyask} with generic $f(t)$. If one takes $f(t)$ of the
special form $t\,\coth(t)$, it is very likely that \eqref{kkssa}
further reduce to the simple form \eqref{rf}. Note that \eqref{rf}
corresponds to replacing the integrals in \eqref{kkssa},
\eqref{kkssb} by one half of the residues of the integrands at
the pole at $t={\ri\pi\over 2}$. Unfortunately, so far we could
not find a way to reduce the integrals to the residues, and thus
\eqref{rf} lacks rigorous proof. But it passes a number of
nontrivial tests, both analytic and numerical. Thus, all
$G_{\pm}^{(s)}$ listed in \eqref{gpmfew} come out identical by
direct evaluation of the integrals from \eqref{kkssa}. For higher
$s$, using \eqref{rf} instead of \eqref{kkssa} dramatically
simplifies calculations, and all analytic expressions for
$G_{\pm}^{(s)}$ listed in Appendix A and beyond in fact depend on
the validity of \eqref{rf}. We take agreement with the numerical
data in Table 3 as further support of \eqref{rf}. On the other
hand, although in deriving the large-$\lambda$ expansion of the
spectrum in Section 4 we have used \eqref{rf}, it is possible to
show that the results for the coefficients $\Phi_{\pm}^{(k)}(l)$
in \eqref{wkbeven}, \eqref{wkbodd} are independent of the validity
of this relation. In particular, all expressions for these
coefficients in Appendix A can be re-derived by a different
(somewhat more complicated) method which does not rely on
\eqref{rf}. Let us stress also that the simplification \eqref{rf}
depends on the special choice $f(t)=t\,\coth(t)$ in
\eqref{lisuyask}. It is unlikely that any simple modification of
\eqref{rf} exists for more general $f(t)$, say of the form
\eqref{sssos}. Therefore, in the analysis of the problem
\eqref{bs0} in the more interesting case of a generic $\alpha$
(which we plan to present in a separate paper), we have to make do
with the integral representation \eqref{kkssa}.

\end{document}